\documentclass[showpacs,preprintnumbers]{revtex4}
\usepackage{amssymb}
\usepackage{amsmath}
\usepackage{graphicx}
\usepackage{dcolumn}
\usepackage{bm}
\usepackage{subfigure}
\usepackage{color}

\begin{document}

\title{Phase transition and entropic force of de Sitter black hole in massive gravity}
\author{Yubo Ma$^{1}$, Yang Zhang$^{1}$, Lichun Zhang$^{1}$, Liang Wu$^{1}$, Ying Gao$^{2}$, Shuo Cao$^{3}$, Yu Pan$^{4*}$}
\address{ $^{1}$Institute of Theoretical Physics, Shanxi Datong University, Datong,
037009, China \\ $^{2}$School of Mathematical and Statistical, Shanxi Datong University, Datong,
037009, China \\ $^{3}$Department of Astronomy, Beijing Normal University, Beijing, 100875, China \\ $^{4}$School of Science, Chongqing University of Posts and Telecommunications, Chongqing, 400065, China}

\thanks{\emph{e-mail:panyu@cqupt.edu.cn}}

\begin{abstract}

It is well known that de Sitter(dS) black holes generally have a black hole horizon and a cosmological horizon, both of which have Hawking radiation. But the radiation temperature of the two horizons is generally different, so dS black holes do not meet the requirements of thermal equilibrium stability, which brings certain difficulties to the study of the thermodynamic characteristics of black holes. In this paper, dS black hole is regarded as a thermodynamic system, and the effective thermodynamic quantities of the system are obtained. The influence of various state parameters on the effective thermodynamic quantities in the massive gravity space-time is discussed. The condition of the phase transition of the de Sitter black hole in massive gravity space-time is given. We consider that the total entropy of the dS black hole is the sum of the corresponding entropy of the two horizons plus an extra term from the correlation of the two horizons. By comparing the entropic force of interaction between black hole horizon and the cosmological horizon with Lennard-Jones force between two particles, we find that the change rule of entropic force between the two system is surprisingly the same. The research will help us to explore the real reason of accelerating expansion of the universe.
\end{abstract}

\maketitle

\section{Introduction}

In recent years, the study of thermodynamic properties of black holes has aroused great interest. Especially, after treating the cosmological constant as the thermodynamic pressure, the first law of black hole thermodynamics in extended phase space has been derived. Comparing the state parameters of black holes with that of the Van der Waals equation, the critical phenomena of various AdS space-time black holes have been studied. The influence of various parameters on the phase transition is discussed. Some progress has been made in the study of black hole phase transition \citep{David Kubiznak12,Robie A. Hennigar17,Aruna Rajagopal14,Hendi17a,Hendi17b,Hendi15,Panahiyan19,Rong-Gen Cai13,Rong-Gen Cai16,Jia-Lin Zhang15a,Jia-Lin Zhang15b,Hendi16,Wei Xu14,Wen14,Wilson G. Brenna15,Rabin Banerjee17,Rabin Banerjee11,Rabin Banerjee12,Meng-Sen Ma17a,Meng-Sen Ma17b,Hendi17c,Dayyani17,De-Cheng Zou17a,Peng Cheng16,Mozhgan Mir17,Zixu Zhao14,Ren Zhao13,Zeinab18,Run Zhou19,Toledo19}. For de Sitter space-time, space-time has black hole horizon and cosmological horizon. Both horizons have Hawking radiation, and the radiation temperatures of the two horizons are generally different. Therefore dS black holes do not meet the requirements of thermal stability, which brings certain difficulties to the study of the thermodynamic properties of black hole. In recent years, the study of the thermodynamic properties of de Sitter space-time has attracted extensive attention \citep{Saoussen Mbarek19,Fil Simovic19,Brian P. Dolan13,Hendi17d,Sekiwa06,David Kubiznak16,James McInerney16,Miho Urano09,Sourav Bhattacharya13,Cai02a,Cai02b,Yu-Bo Ma18a,Azreg-Aiou15,Zhang16,Li-Chun Zhang14,Hui-Hua Zhao14,Huaifan Li17,Li-Chun Zhang16,Yu-Bo Ma18b}. Because in the early period of inflation, our universe is a quasi de Sitter space-time, and the constant term introduced in the study of de Sitter space-time is the contribution of vacuum energy, which is also a form of material energy. If the cosmological constant corresponds to dark energy, our universe will evolve into a new de Sitter phase. In order to construct the whole history of universe evolution, we should have a clear understanding of the classical and quantum properties of the de Sitter space-time.

Since the thermodynamic quantities corresponding to the two horizons in de Sitter space-time are all functions of mass $M$, charge $Q$ and cosmological constant $\Lambda$. There must be certain relation between the thermodynamic quantities corresponding to the two horizons. Considering whether the corresponding thermodynamic quantity in the de Sitter space-time has the thermodynamic characteristics of AdS black hole after the correlation of black hole horizon and universe horizon; What is the relationship between the entropy of de Sitter space and the corresponding entropy of two horizons? What is the relationship between the entropic force caused by the interaction of two horizons and the ratio of the position of the two horizons? These problems are very important to study the stability and evolution of de Sitter space-time. Therefore, it is worth to establish a self-consistent relationship between the thermodynamic quantities of de Sitter space-time. In \citep{Hinterbichler12}, the authors suggest a massive gravity which overcomes the traditional problems and yield an avenue for addressing the cosmological constant naturalness problem. Fierz and Pauli firstly provided the possibility of a massive graviton \citep{Fierz1939a,Fierz1939b}. Further,  van Dam and Veltman \citep{van Dam1970} and Zakharov \citep{Zakharov1970} had found the linear theory which was coupled to a source, they found that even when the mass of the graviton approaches to zero, the prediction of the theory is different from that of the linear theory. Later, Boulware-Deser ghost theories had been studied, which were nonlinear and ghostlike instability massive gravity theories \citep{Boulware1972a,Boulware1972b}. In the absence of such instability, significant progress has been made in establishing the theory of massive gravity \citep{de Rham10,de Rham11,B. Eslam Panah19b,B. Eslam Panah19a,B. Eslam Panah20,S. H. Hendi17c}. The most straightforward way to construct the massive gravity theories is to simply add a mass term to the GR action, giving the graviton a mass in such a way that GR is recovered as mass vanishes. Recently, a charged BTZ black holes has been studied in \citep{Hendi16,Hendi17f}, which consider a massive gravity. The massive BTZ black holes in the presence of Maxwell and Born-Infeld electrodynamics in asymptotically (A)dS spacetimes was studied in \citep{Hendi15}. In \citep{Majhi17,Cai15,Jianfei Xu15,Upadhyay}, thermodynamics and entropy of AdS black hole in massive gravity were obtained, and the phase transition was discussed.

In this paper, on the basis of effective thermodynamic quantity of de Sitter black hole in massive gravity (DSBHMG), the thermodynamic characteristics of de Sitter space-time are discussed. The condition of the phase transition of DSBHMG space-time is found, and the influence of each parameter on the critical point is analyzed. It is noted that DSBHMG space-time has a second-order phase transition similar to that AdS black hole when the parameters meet certain conditions, except that the heat capacity at constant volume of DSBHMG  is not zero. By calculating the entropic force of the interaction between the two horizons, the changing law of entropic force with the ratio of the position of two horizons is given. By comparing the curve with that of Lennard-Jones force \citep{David C. Johnston} between two particles, it is found that the curve of entropic force changing with the position ratio of two horizons is similar to the curve of Lennard-Jones force changing with the distance between two particles, which has the same change rule. This discovery provides a new way to study the internal cause of accelerating expansion of the universe.

This paper is arranged as follows: for the continuity of this paper, in the second part, we briefly introduce the corresponding thermodynamic quantities of black hole horizon and cosmological horizon in DSBHMG and the effective thermodynamic quantities of DSBHMG. In the third part, we discuss the critical phenomena of DSBHMG. In the fourth part, we discuss the entropic force of interaction between the two horizons, and obtain the entropic force and the Lennard-Jones force between two particles has a similar law of change. The fifth part of the paper discusses the conclusion. (We use the units $G=h=k_B=1$).

\section{Thermodynamics of black holes in massive gravity}

\label{sec:method}
 We consider $(3+1)$ -dimensional massive gravity with a Maxwell field \citep{Cai15,Jianfei Xu15,Upadhyay,Hassan11,Allan Adams15}, the action is as follows
\begin{equation}\label{2.1}
S=\frac{1}{k^2}\int d^4x\sqrt{g}(R-2\Lambda-\frac{1}{4}F^2+m^2\sum_i^4{c_i\mu_i}),
\end{equation}
where $\Lambda$ is the cosmological constant and $k=0$,$1$, or $-1$ , correspond to a Ricci flat, sphere, or hyperbolic horizon for the black hole, respectively. Here  $F_{\mu\nu}$ is the Maxwell field-strength tensor, $c_i$ are constants, and $\mu_i$ are symmetric polynomials of the eigenvalues of the matrix $\sqrt{g^{\mu\nu}f_{\mu\nu}} $ where $f_{\mu\nu}$ is a fixed symmetric tensor.
The action admits a static black hole solution with the space-time metric and reference metric as

\begin{equation}\label{2.2}
d s^{2}=-f(r) d t^{2}+f^{-1} d r^{2}+r^{2} h_{ij} d x^{i} d x^{j}, \quad (i, j=1,2)
\end{equation}
where $h_{ij}dx^{i}dx^{j}$ is the line element for an Einstein space with constant curvature, $f(r)$ is the metric function, which is written by \citep{De-Cheng Zou17b,Petarpa Boonserm18}.

\begin{equation}\label{2.3}
f(r)=k-\frac{\Lambda}{3} r^{2}-\frac{m_{0}}{r}+\frac{q^{2}}{4 r^{2}}+\frac{c_{1} m^{2}}{2} r+m^{2} c_{2},
\end{equation}

The positions of black hole horizon and universe horizon satisfy the equation $f(r_{+,c})=0$, thus, the mass $m_0$ can be expressed in terms of $r_{+,c}$ as
\begin{equation}\label{2.4}
M=\frac{m_{0}}{2}=\frac{\left(k+m^{2} c_{2}\right) r_{c} x(1+x)}{2\left(1+x+x^{2}\right)}+\frac{q^{2}(1+x)\left(1+x^{2}\right)}{8 r_{c} x\left(1+x+x^{2}\right)}+\frac{r_{c}^{2} m^{2} c_{1} x^{2}}{4\left(1+x+x^{2}\right)},
\end{equation}

where $x=\frac{r_+}{r_c}$ and

\begin{equation}\label{2.5}
\frac{\Lambda}{3} r_{c}^{2}\left(1+x+x^{2}\right)=k-\frac{q^{2}}{4 r_{c}^{2} x}+\frac{c_{1}}{2} m^{2} r_{c}(1+x)+m^{2} c_{2},
\end{equation}

The radiation temperature of black hole horizon and cosmological horizon is

\begin{equation}\label{2.6}
T_{+, c}=\pm \frac{f\left(r_{+, c}\right)}{4 \pi}=\frac{1}{4 \pi r_{+, c}}\left(k-\Lambda r_{+, c}^{2}-\frac{q^{2}}{4 r_{+, c}^{2}}+m^{2} c_{1} r_{+, c}+m^{2} c_{2}\right).
\end{equation}

The thermodynamic quantities corresponding to the two horizons satisfy the first law of thermodynamics

\begin{equation}\label{2.6.1}
d M=T_{+, c} d S_{+, c}+V_{+, c} d P+\mu_{+, c} d Q,
\end{equation}

where

\begin{equation}\label{2.7}
V_{+, c}=\frac{v_{2}}{3} r_{+, c}^{3}, \quad P=-\frac{\Lambda}{8 \pi}
\end{equation}
when the radiation temperature $T_{+}$ of the black hole horizon is equal to that of the cosmological horizon $T_{c}$, the charge $Q$ and $\Lambda$ of the system satisfies the equation

\begin{equation}\label{2.8}
\frac{1}{r_{+}}\left(k-\Lambda r_{+}^{2}-\frac{q^{2}}{4 r_{+}^{2}}+m^{2} c_{1} r_{+}+m^{2} c_{2}\right)=-\frac{1}{r_{c}}\left(k-\Lambda r_{c}^{2}-\frac{q^{2}}{4 r_{c}^{2}}+m^{2} c_{1} r_{c}+m^{2} c_{2}\right),
\end{equation}

we can get the radiation temperature $T$ when $T_+ = T_c$.

\begin{equation}\label{2.9}
T=T_{+}=T_{c}=\frac{(1-x)}{2 \pi r_{c}(1+x)^{2}}\left[k+\frac{m^{2} c_{1} r_{c}\left(1+4 x+x^{2}\right)}{4(1+x)}+m^{2} c_{2}\right].
\end{equation}

Recently, through the study of the thermodynamic characteristics of dS space-time, one can get the thermodynamic volume of dS space-time is \citep{Brian P. Dolan13,Li-Chun Zhang16}

\begin{equation}\label{2.10}
V=\frac{4 \pi}{3}\left(r_{c}^{3}-r_{+}^{3}\right)=\frac{4 \pi}{3} r_{c}^{3}\left(1-x^{3}\right),
\end{equation}
if the energy $M$, charge $Q$ and volume $V$ in space-time are taken as the state parameters of a thermodynamic system, the first law of thermodynamics should be satisfied

\begin{equation}\label{2.11}
d M=T_{e f f} d S-P_{e f f} d V+\phi_{eff} d Q,
\end{equation}

Considering the interaction between the black hole horizon and the cosmological horizon, we set the entropy of the system \citep{Huaifan Li17,Li-Chun Zhang16,Yu-Bo Ma18b}

\begin{equation}\label{2.12}
S=\pi r_{c}^{2}\left[1+x^{2}+f(x)\right],
\end{equation}
here the undefined function $f(x)$ represents the extra contribution from the correlations of the two horizons.
When the radiation temperature of black hole horizon and cosmological horizon is equal, the effective temperature of space-time should also be equal to the radiation temperature of two horizons

\begin{equation}\label{2.13}
T_{e f f}=T=T_{c}=T_{+},
\end{equation}

From equations (\ref{2.11}), (\ref{2.12}) and (\ref{2.13}), the effective temperature $T_{eff}$, pressure $P_{eff}$ and potential $\phi_{eff}$ of the system are obtained

\begin{equation}\label{2.14}
T_{e f f}=\frac{B(x, q)\left(1-x^{3}\right)}{4 \pi r_{c} x\left(1+x^{4}\right)}, \quad P_{e f f}=\frac{D(x, q)\left(1-x^{3}\right)}{16 \pi r_{c}^{2} x\left(1+x^{4}\right)},
\quad \phi_{e f f}=\left(\frac{\partial M}{\partial Q}\right)_{S, V}=\frac{4 \pi^{2} Q(1+x)\left(1+x^{2}\right)}{v_{2}^{2} r_{c} x\left(1+x+x^{2}\right)}.
\end{equation}

where
\begin{equation}\label{2.15}
\begin{aligned}
B(x, q)=&\left(k+m^{2} c_{2}\right) \frac{\left(1+x-2 x^{2}+x^{3}+x^{4}\right)}{\left(1+x+x^{2}\right)}\\
&-\frac{q^{2}}{4 r_{c}^{2}}\frac{\left(1+x+x^{2}-2x^{3}+x^{4}+x^{5}+x^{6}\right)}{x^{2}\left(1+x+x^{2}\right)}
+r_{c} m^{2} c_{1} x \frac{(2+x)(1-x)+2 x^{3}}{2\left(1+x+x^{2}\right)},
\end{aligned}
\end{equation}

\begin{equation}\label{2.16}
\begin{aligned}
D(x, q)=&\frac{\left(k+m^{2} c_{2}\right)}{\left(1+x+x^{2}\right)^{2}} \left\{(1+2x)\left[1+x^{2}+f(x)\right]-x(1+x)\left(1+x+x^{2}\right)[2 x+f'(x)]\right\} \\
&-\frac{q^{2}}{4 r_{c}^{2} x^{2}\left(1+x+x^{2}\right)^{2}}\left\{2\left(1+2x+3x^{2}\right)\left[1+x^{2}+f(x)\right]-x(1+x)\left(1+x^{2}\right)\left(1+x+x^{2}\right)[2 x+f(x)]\right\} \\
&+\frac{r_{c} m^{2} c_{1} x}{\left(1+x+x^{2}\right)^{2}}\left\{2(2+x)\left[1+x^{2}+f(x)\right]-x\left(1+x+x^{2}\right)\left[2 x+f^{\prime}(x)\right]\right\}.
\end{aligned}
\end{equation}

\begin{figure}[htp]

\centering
\subfigure[]{\includegraphics[width=0.30\textwidth]{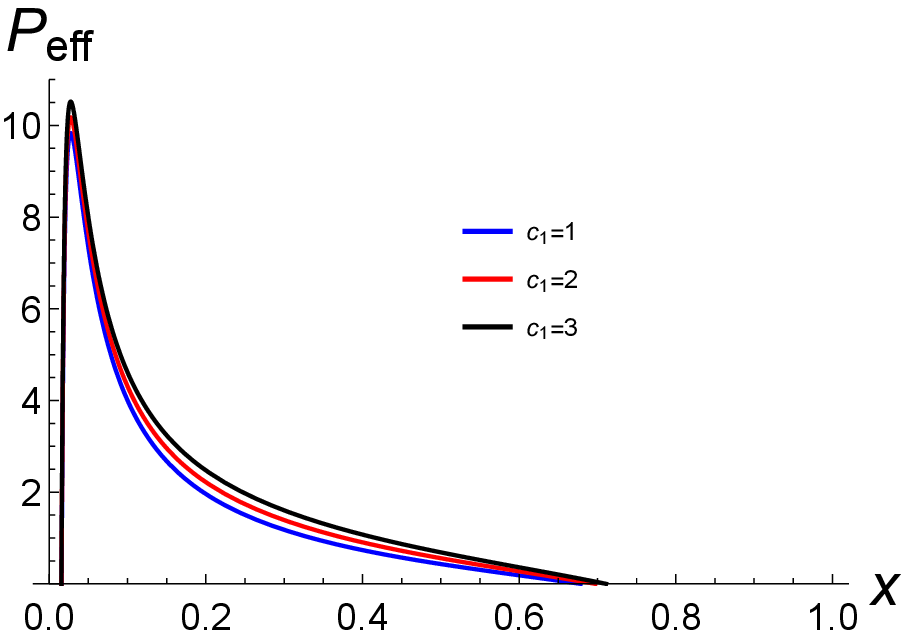}}
\subfigure[]{\includegraphics[width=0.30\textwidth]{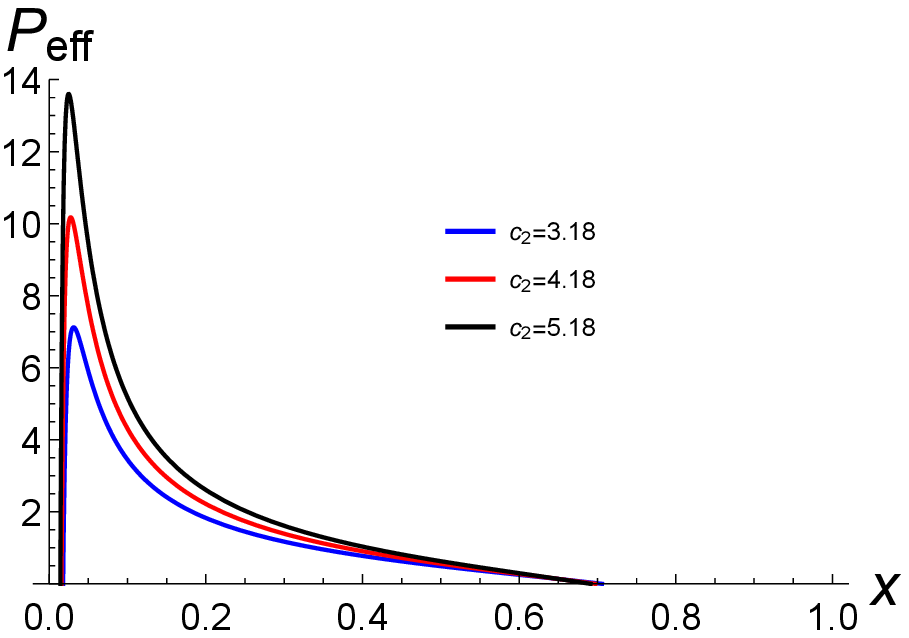}}\newline
\subfigure[]{\includegraphics[width=0.30\textwidth]{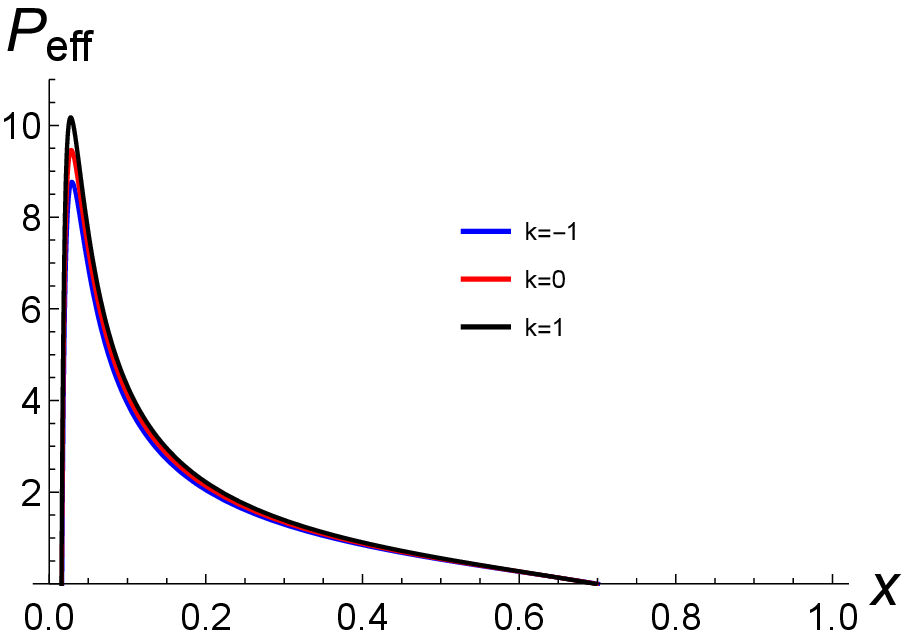}}
\subfigure[]{\includegraphics[width=0.30\textwidth]{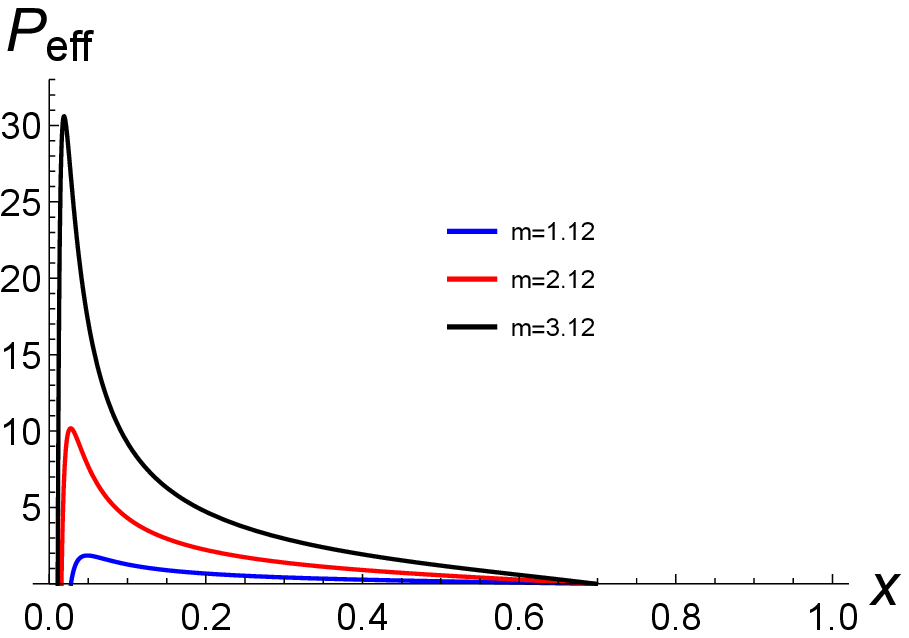}}
\subfigure[]{\includegraphics[width=0.30\textwidth]{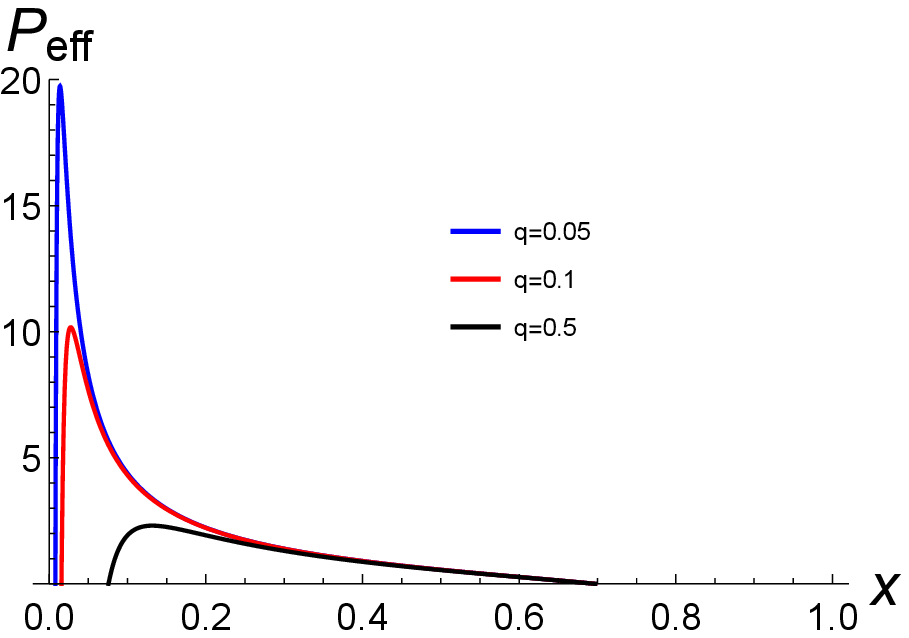}}\newline
\caption{$P_{eff}-x$ diagrams when the parameters change respectively.}
\label{fig1}
\end{figure}

\begin{figure}[htp]
\centering
\subfigure[]{\includegraphics[width=0.30\textwidth]{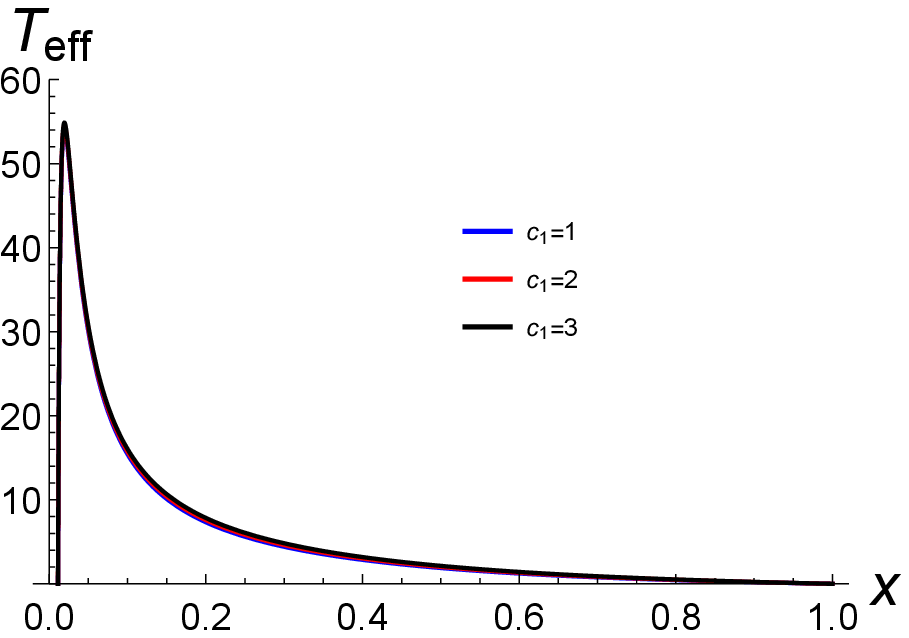}}
\subfigure[]{\includegraphics[width=0.30\textwidth]{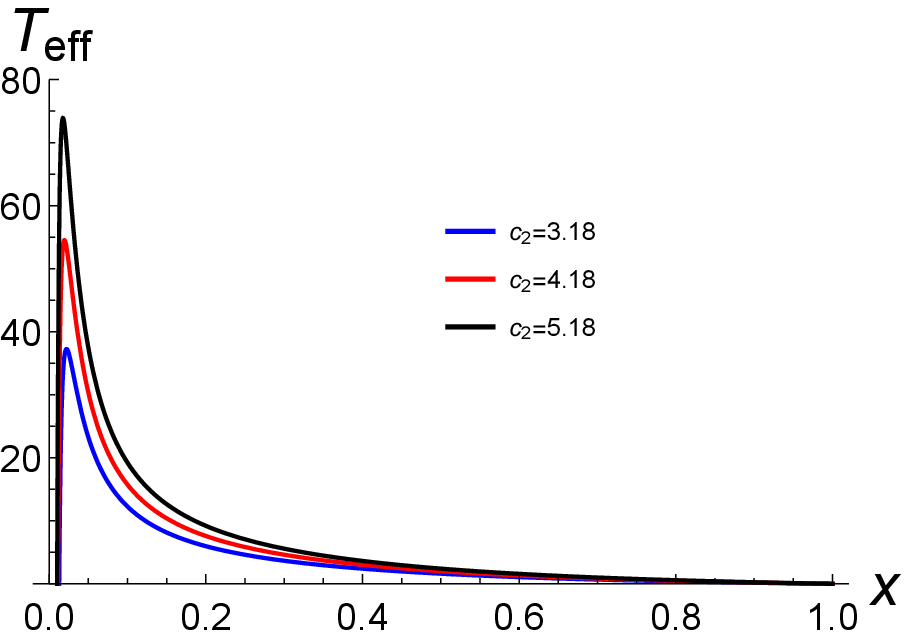}}\newline
\subfigure[]{\includegraphics[width=0.30\textwidth]{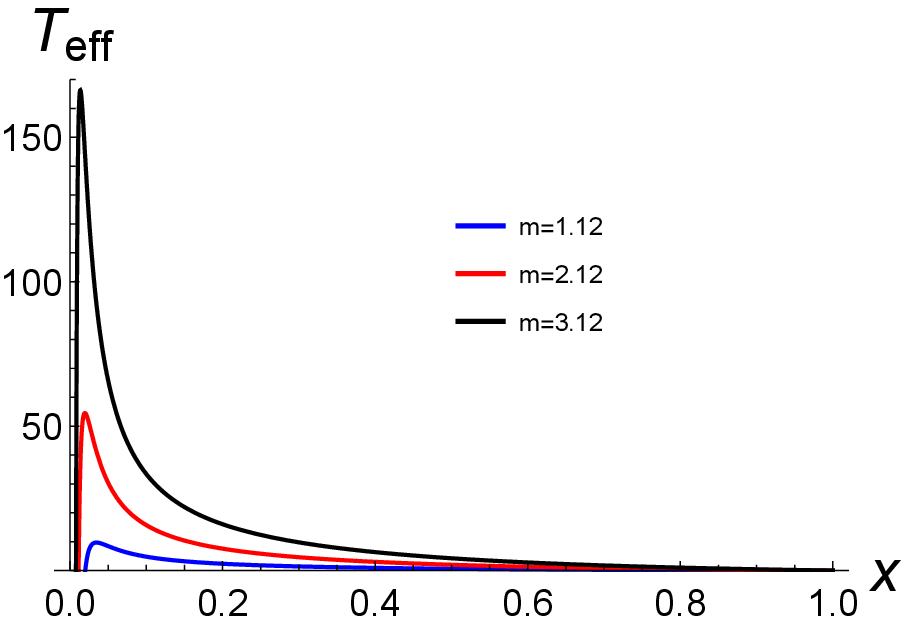}}
\subfigure[]{\includegraphics[width=0.30\textwidth]{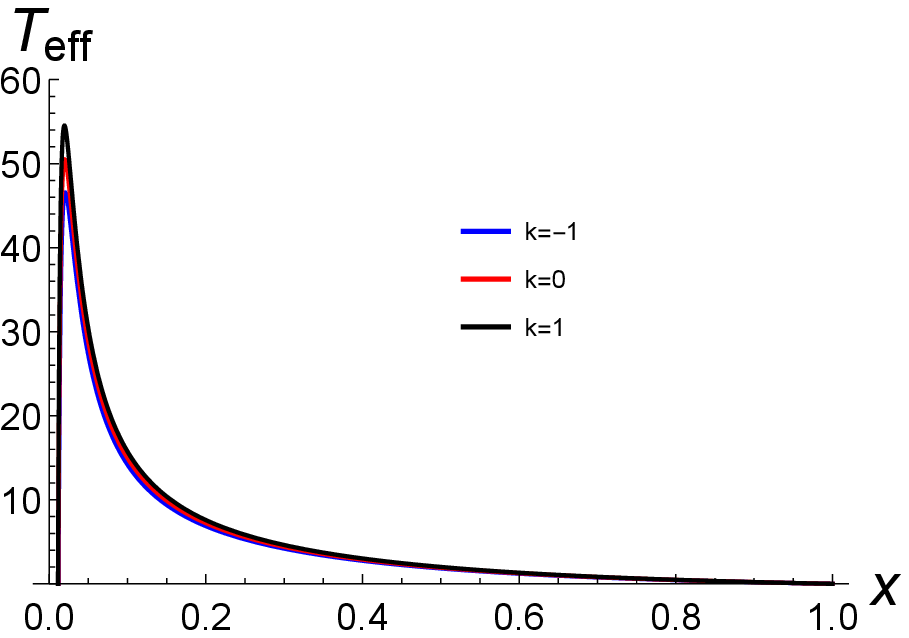}}
\subfigure[]{\includegraphics[width=0.30\textwidth]{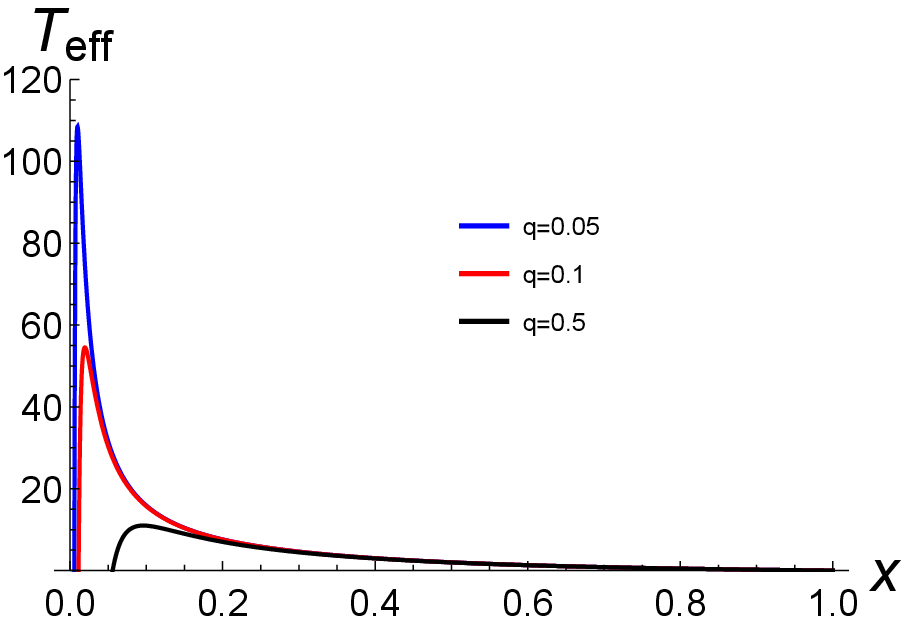}}
\caption{$T_{eff}-x$ diagrams when the parameters change respectively.}\label{fig2}
\end{figure}

Considering the initial conditions $f(0)=0$, we can get

\begin{equation}\label{2.17}
f(x)=-\frac{2\left(4-5 x^{3}-x^{5}\right)}{5\left(1-x^{3}\right)}+\frac{8}{5}\left(1-x^{3}\right)^{2 / 3}
\end{equation}
we set the initial parameters to $m=2.12$, $c_1=2$,$c_2=3.18$,$q=1.7$, $r_c=1$,$k=1$, and then take different values for $k$, $q$, $m$, $c_1$ and $c_2$ respectively, to get the curve $P_{eff}-x$ and $T_{eff}-x$ curve (take $r_c=1$) as shown in FIG.\ref{fig1} and FIG.\ref{fig2}

\section{Critical phenomena}

Heat capacity of the system at constant volume:
\begin{equation}\label{3.1}
\begin{aligned}
C_{V}&=T_{eff}\left(\frac{\partial S}{\partial T_{e f f}}\right)_{V}=T_{eff} \frac{\left(\frac{\partial S}{\partial r_{c}}\right)_{x}\left(\frac{\partial V}{\partial x}\right)_{r_{c}}-\left(\frac{\partial S}{\partial x}\right)_{r_{c}}\left(\frac{\partial V}{\partial r_{c}}\right)_{x}}{\left(\frac{\partial V}{\partial x}\right)_{r_{c}}\left(\frac{\partial T_{eff}}{\partial_{r_{c}}}\right)_{x}-\left(\frac{\partial V}{\partial r_{c}}\right)_{x}\left(\frac{\partial T_{eff}}{\partial x}\right)_{r_{c}}} \\
&=\frac{2 \pi r_{c}^{2} B(x, q) x\left(1+x^{4}\right)}{\left(1-x^{3}\right)\left[B'(x, q)(1-x^{3}) x^{2}\bar{B}(x, q)-\frac{B(x, q)(1+2 x^{3}+5 x^{4}-2 x^{7})}{x\left(1+x^{4}\right)}\right]}
\end{aligned}
\end{equation}
where
\begin{equation}\label{3.2}
\bar{B}(x, q)=\left(k+m^{2} c_{2}\right) \frac{\left(1+x-2 x^{2}+x^{3}+x^{4}\right)}{\left(1+x+x^{2}\right)}-\frac{3}{4} \mu_{c}^{2} \frac{\left(1+x+x^{2}-2 x^{3}+x^{4}+x^{5}+x^{6}\right)}{x^{2}\left(1+x+x^{2}\right)}
\end{equation}
$B'(x,q)=\frac{d B(x,q)}{d x}$, and $B(x,q)$ is presented by Eq.(\ref{2.15}).

\begin{figure}[htp]
\centering
\subfigure[]{\includegraphics[width=0.30\textwidth]{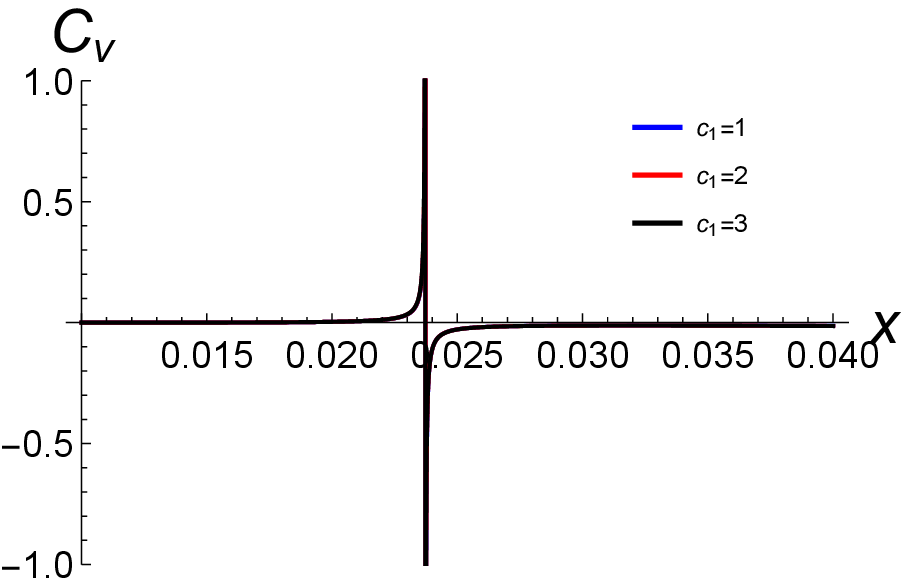}}
\subfigure[]{\includegraphics[width=0.30\textwidth]{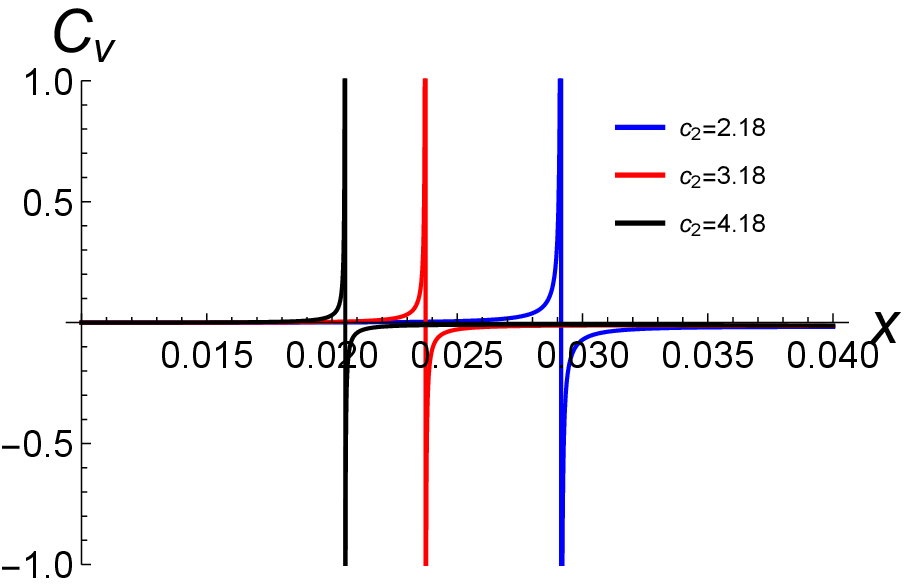}}\newline
\subfigure[]{\includegraphics[width=0.30\textwidth]{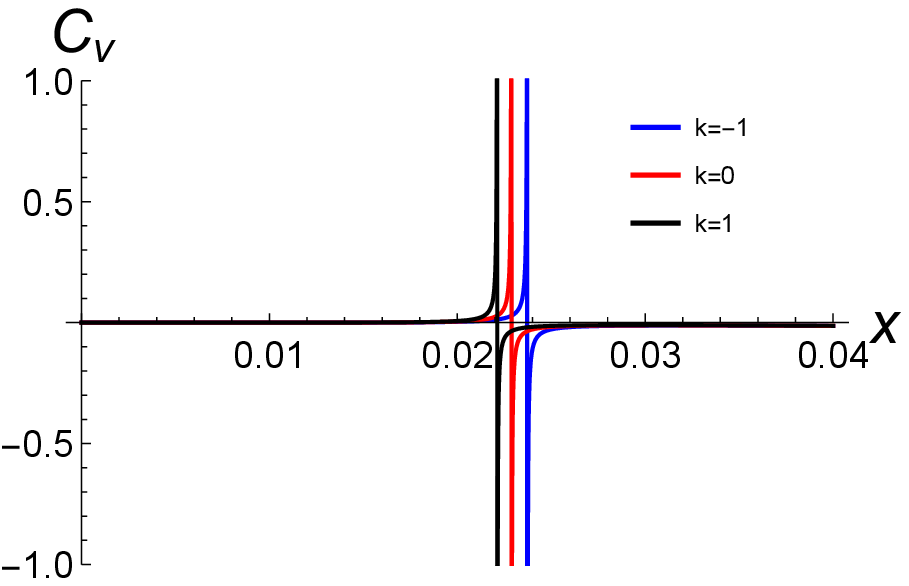}}
\subfigure[]{\includegraphics[width=0.30\textwidth]{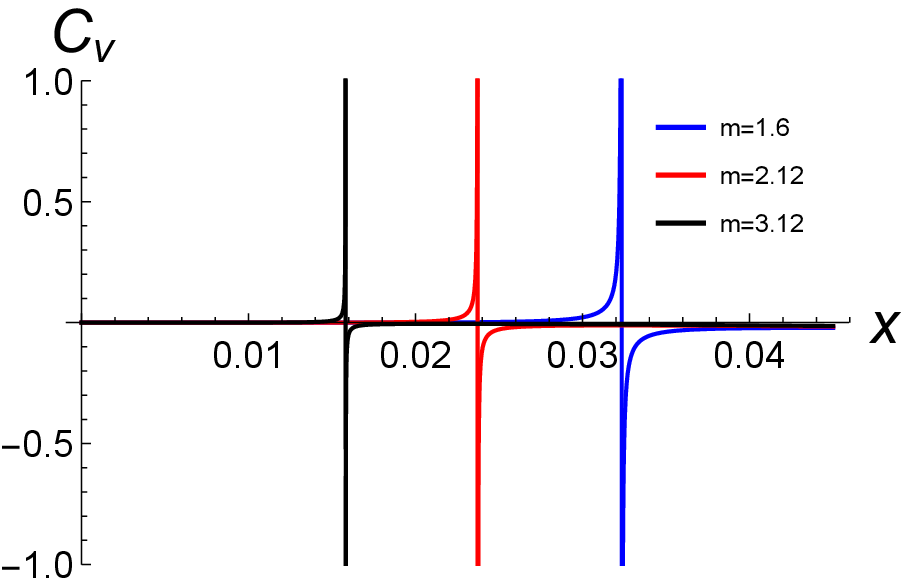}}
\subfigure[]{\includegraphics[width=0.30\textwidth]{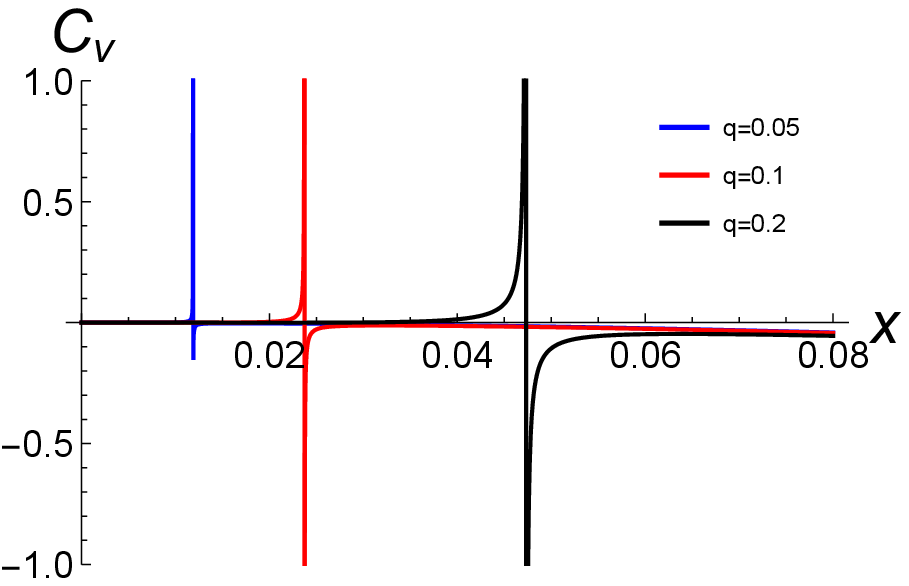}}
\caption{  $C_V-x$ diagrams when the parameters change respectively and the initial parameters  $m=2.12$, $c_1=2$,$c_2=3.18$,$q=1.7$, $r_c=1$,$k=1$.}
\label{fig3}
\end{figure}

From FIG.\ref{fig3}, we know that the heat capacity $C_V$ of the system in DSBHMG is not zero, which is different from the result for AdS black holes where $C_V=0$ .

The heat capacity at constant pressure is

 \begin{equation}\label{3.3}
 C_{P_{eff}}=T_{e f f}\left(\frac{\partial S}{\partial T_{eff}}\right)_{P_{eff}}=T_{eff} \frac{\left(\frac{\partial S}{\partial r_{c}}\right)_{x}\left(\frac{\partial P_{eff}}{\partial x}\right)_{r_{c}}-\left(\frac{\partial S}{\partial x}\right)_{r_{c}}\left(\frac{\partial P_{eff}}{\partial r_{c}}\right)_{x}}{\left(\frac{\partial T_{eff}}{\partial r_{c}}\right)_{x}-\left(\frac{\partial P_{eff}}{\partial r_{c}}\right)_{x}\left(\frac{\partial T_{eff}}{\partial x}\right)_{r_{c}}}=\frac{2 \pi r_{c}^{2} B(x, q) E(x, q)}{F(x, q)}
\end{equation}
Here
 \begin{equation}\label{3.4.1}
 E(x, q)=\left[1+x^{2}+f(x)\right]\left[D^{\prime}(x, q)\left(1-x^{3}\right)-\frac{D(x, q)\left(1+2 x^{3}+5 x^{4}-2 x^{7}\right)}{x\left(1+x^{4}\right)}\right]+\left[2 x+f^{\prime}(x)\right] \frac{\bar{D}(x, q)\left(1-x^{3}\right)}{2 x},
\end{equation}

\begin{equation}\label{3.4.2}
\begin{aligned}
\bar{D}(x, q)=&\frac{2\left(k+m^{2} c_{2}\right)}{\left(1+x+x^{2}\right)^{2}}\left\{(1+2 x)\left[1+x^{2}+f(x)\right]-x(1+x)\left(1+x+x^{2}\right)\left[2 x+f'(x)\right]\right\} \\
&-\frac{\mu_{c}^{2}}{x^{2}\left(1+x+x^{2}\right)^{2}}\left\{2\left(1+2x+3x^{2}\right)\left[1+x^{2}+f(x)\right]-x(1+x)\left(1+x^{2}\right)\left(1+x+x^{2}\right)\left[2 x+f^{\prime}(x)\right]\right\} \\
&{+\frac{r_{c} m^{2} c_{1} x}{\left(1+x+x^{2}\right)^{2}}\left\{2(2+x)\left[1+x^{2}+f(x)\right]-x\left(1+x+x^{2}\right)\left[2 x+f^{\prime}(x)\right]\right\}},
\end{aligned}
\end{equation}

\begin{equation}\label{3.4.3}
\begin{aligned}
F(x, q)=&\bar{B}(x, q)\left[\frac{D(x, q)\left(1+2 x^{3}+5 x^{4}-2 x^{7}\right)}{x\left(1+x^{4}\right)}-D'(x,q)\left(1-x^{3}\right)\right] \\
&+\bar{D}(x, q)\left[B^{\prime}(x, q)\left(1-x^{3}\right)-\frac{B(x, q)\left(1+2 x^{3}+5 x^{4}-2 x^{7}\right)}{x\left(1+x^{4}\right)}\right].
\end{aligned}
\end{equation}

\begin{figure}[htp]
\centering
\subfigure[]{\includegraphics[width=0.30\textwidth]{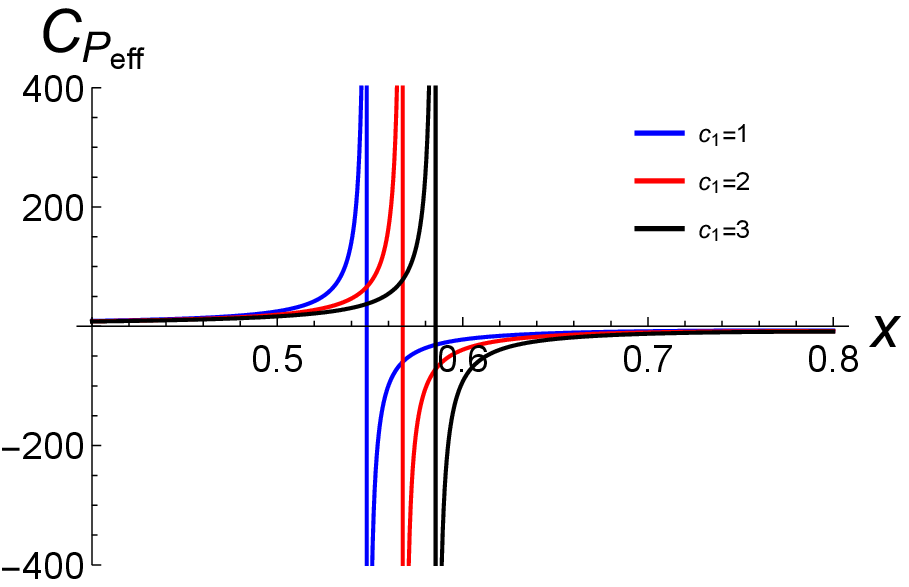}}
\subfigure[]{\includegraphics[width=0.30\textwidth]{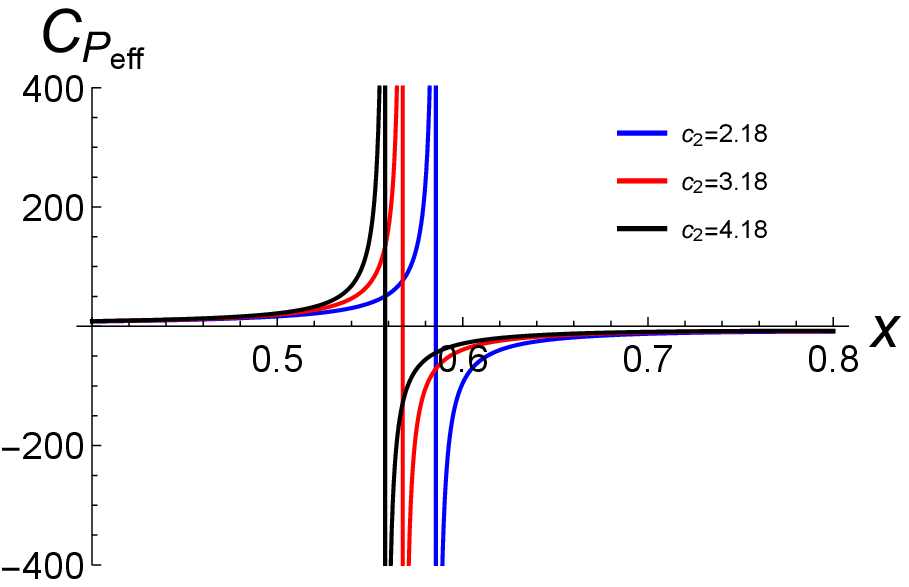}}\newline
\subfigure[]{\includegraphics[width=0.30\textwidth]{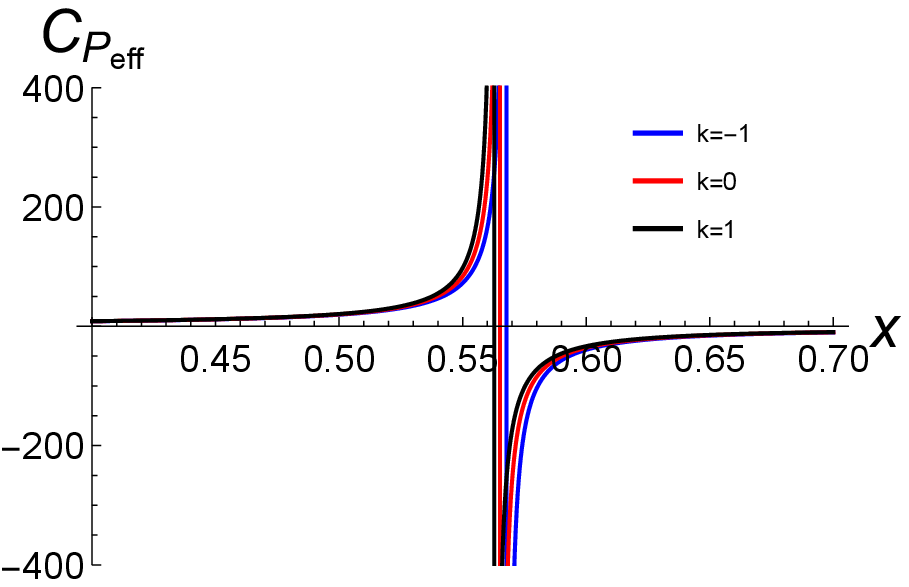}}
\subfigure[]{\includegraphics[width=0.30\textwidth]{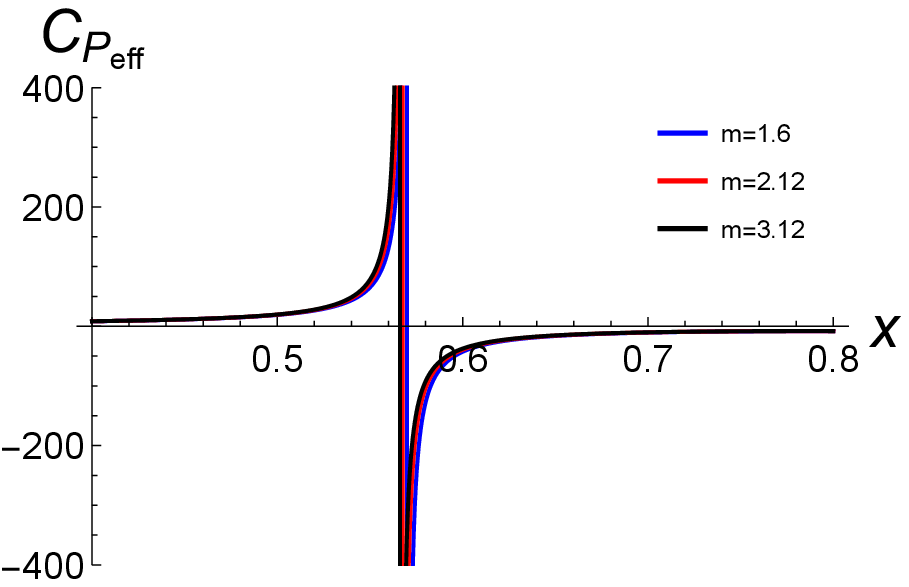}}
\subfigure[]{\includegraphics[width=0.30\textwidth]{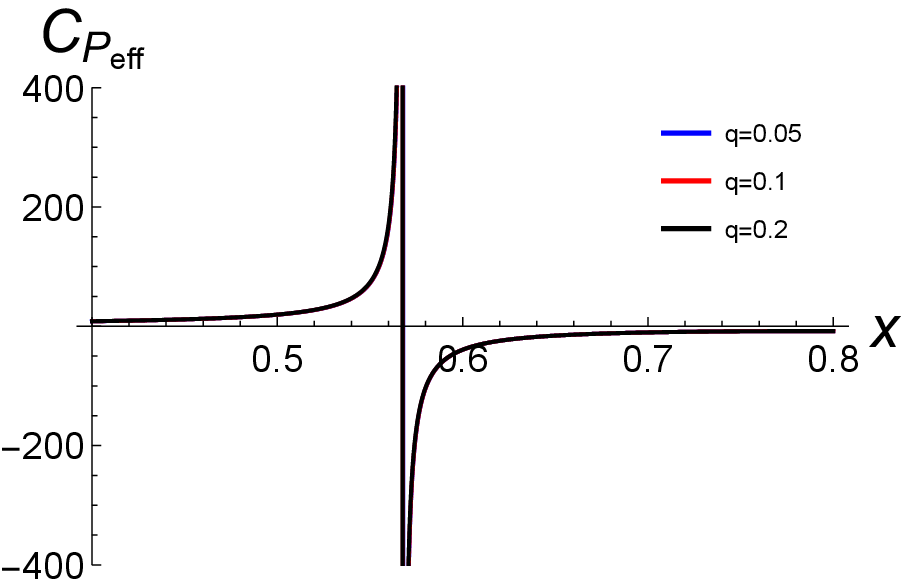}}
\caption{$C_{P_{eff}}-x$ diagrams when the parameters change respectively and the initial parameters  $m=2.12$, $c_1=2$,$c_2=3.18$,$q=1.7$, $r_c=1$,$k=1$.}
\label{fig4}
\end{figure}

$D'(x,q)=\frac{xD(x,q)}{dx}$, and $D(x,q)$ is given by Eq.(\ref{2.16}).From Eq.(\ref{3.3}), the curves are depicted in FIG.\ref{fig4}.

The expansion coefficient is
\begin{equation}\label{3.5}
\begin{aligned}
\alpha &=\frac{1}{V}\left(\frac{\partial V}{\partial T_{eff}}\right)_{P_{eff}}\\
&=\frac{1}{V} \frac{\left(\frac{\partial V}{\partial r_{c}}\right)_{x}\left(\frac{\partial P_{eff}}{\partial x}\right)_{r_{c}}-\left(\frac{\partial V}{\partial x}\right)_{r_{c}}\left(\frac{\partial P_{eff}}{\partial r_{c}}\right)_{x}}{\left(\frac{\partial P_{eff}}{\partial x}\right)_{r_{c}}\left(\frac{\partial T_{eff}}{\partial r_{c}}\right)_{x}-\left(\frac{\partial P_{eff}}{\partial r_{c}}\right)_{x}\left(\frac{\partial T_{eff}}{\partial x}\right)_{r_{c}}} \\
&= \frac{12 r_{c} x\left(1+x^{4}\right)}{\left(1-x^{3}\right) F(x, q)}\left[D'(x, q)\left(1-x^{3}\right)-\frac{D(x, q)\left(1+2 x^{3}+5 x^{4}-2 x^{7}\right)}{x\left(1+x^{4}\right)}-\bar{D}(x, q) x^{2}\right]
\end{aligned}
\end{equation}

\begin{figure}[htp]
\centering
\subfigure[]{\includegraphics[width=0.30\textwidth]{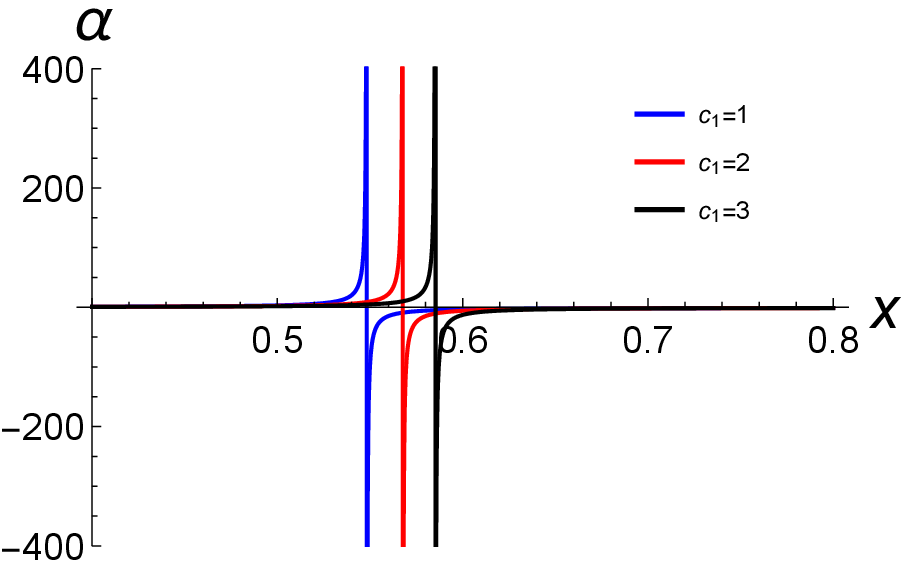}}
\subfigure[]{\includegraphics[width=0.30\textwidth]{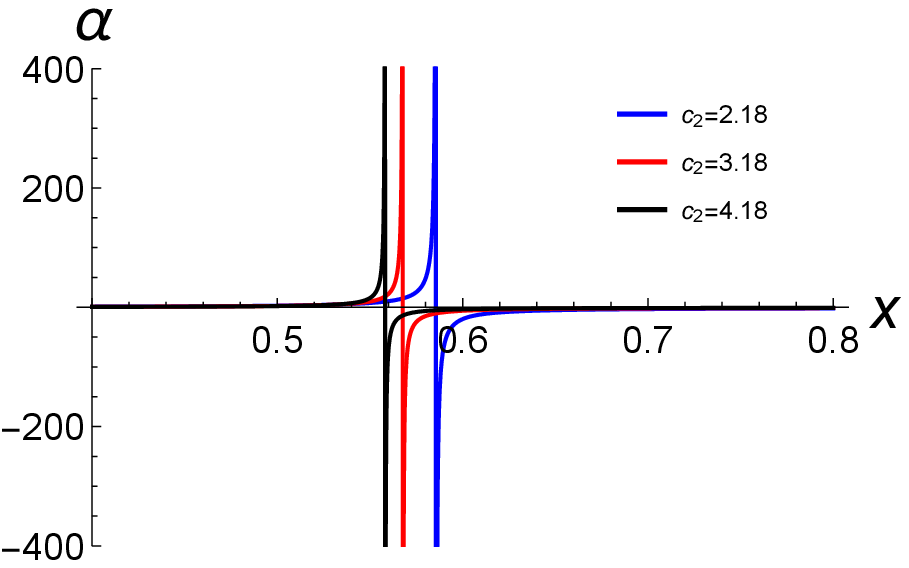}}\newline
\subfigure[]{\includegraphics[width=0.30\textwidth]{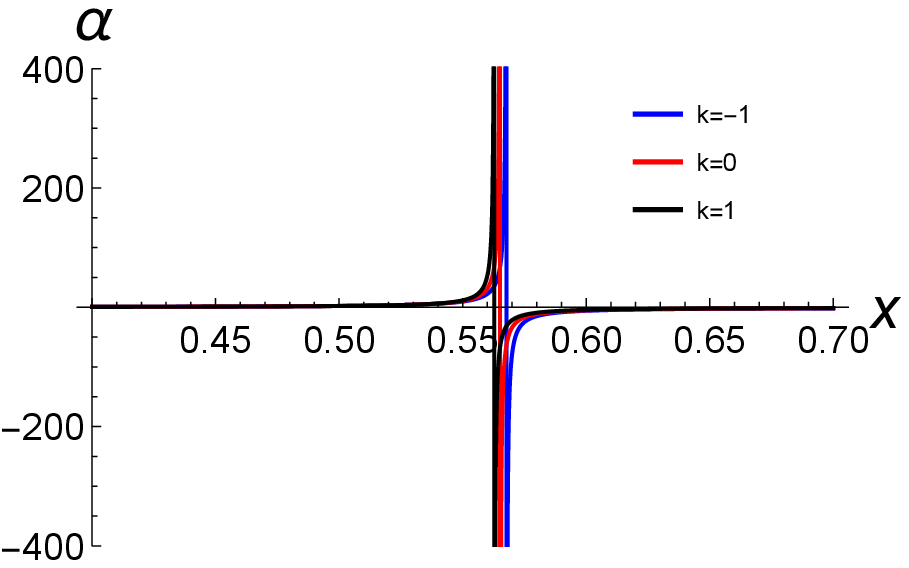}}
\subfigure[]{\includegraphics[width=0.30\textwidth]{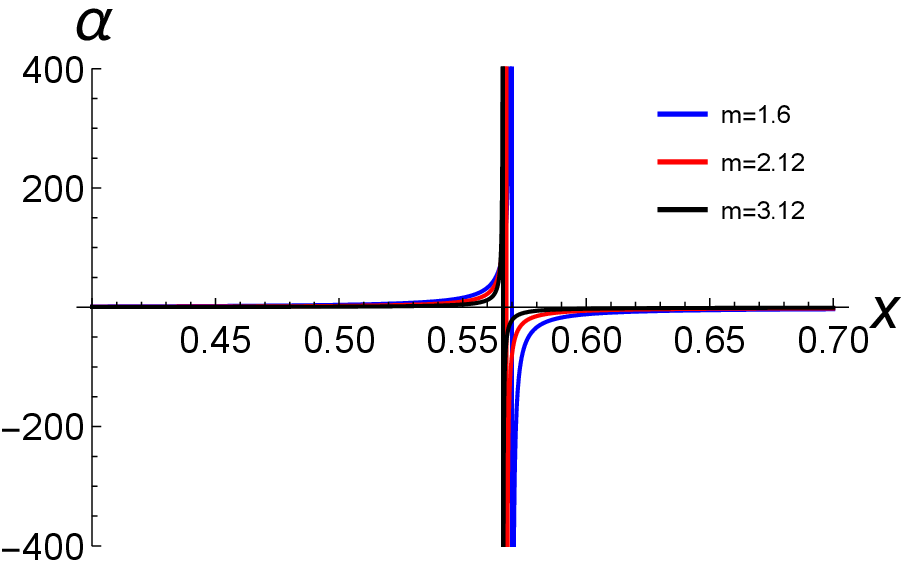}}
\subfigure[]{\includegraphics[width=0.30\textwidth]{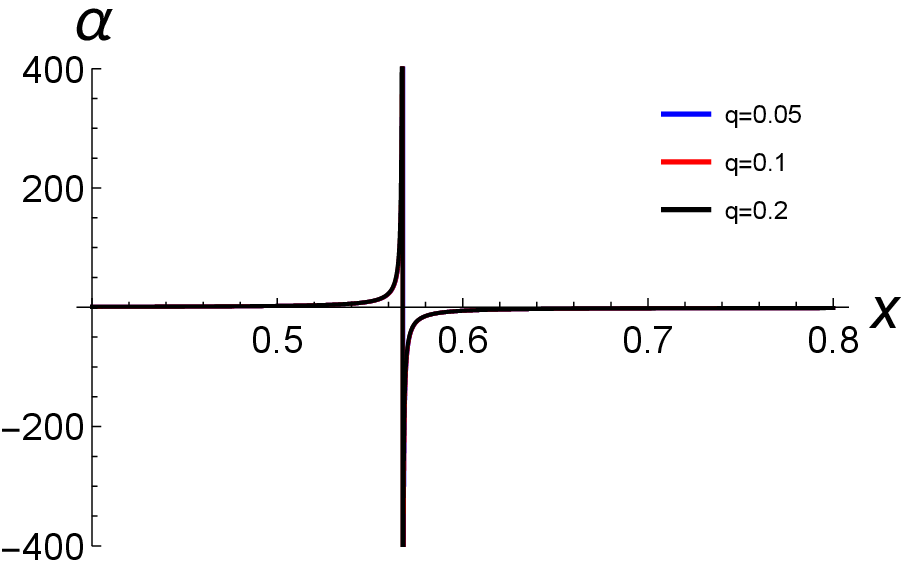}}
\caption{$\alpha-x$ diagrams when the parameters change respectively and the initial parameters  $m=2.12$, $c_1=2$,$c_2=3.18$,$q=1.7$, $r_c=1$,$k=1$.}
\label{fig5}
\end{figure}

The $\alpha-x$ curve is shown in FIG.\ref{fig5}.

The isothermal compressibility is
\begin{equation}\label{3.6}
\begin{aligned}
\kappa_{T_{eff}}&=-\frac{1}{V}\left(\frac{\partial V}{\partial P_{eff}}\right)_{T_{eff}}\\
&=\frac{1}{V} \frac{\left(\frac{\partial V}{\partial r_{c}}\right)_{x}\left(\frac{\partial T_{eff}}{\partial x}\right)_{r_{c}}-\left(\frac{\partial V}{\partial x}\right)_{r_{c}}\left(\frac{\partial T_{eff}}{\partial r_{c}}\right)_{x}}{\left(\frac{\partial P_{eff}}{\partial x}\right)_{r_{c}}\left(\frac{\partial T_{eff}}{\partial r_{c}}\right)_{x}-\left(\frac{\partial P_{eff}}{\partial r_{c}}\right)_{x}\left(\frac{\partial T_{eff}}{\partial x}\right)_{r_{c}}}\\
&=\frac{48 \pi r_{c}^{2} x\left(1+x^{4}\right)}{\left(1-x^{3}\right) F(x, q)}\left[B'(x, q)\left(1-x^{3}\right)-\frac{B(x, q)\left(1+2 x^{3}+5 x^{4}-2 x^{7}\right)}{x\left(1+x^{4}\right)}-\bar{B}(x, q) x^{2}\right].
\end{aligned}
\end{equation}

\begin{figure}[htp]
\centering
 \subfigure[]{\includegraphics[width=0.30\textwidth]{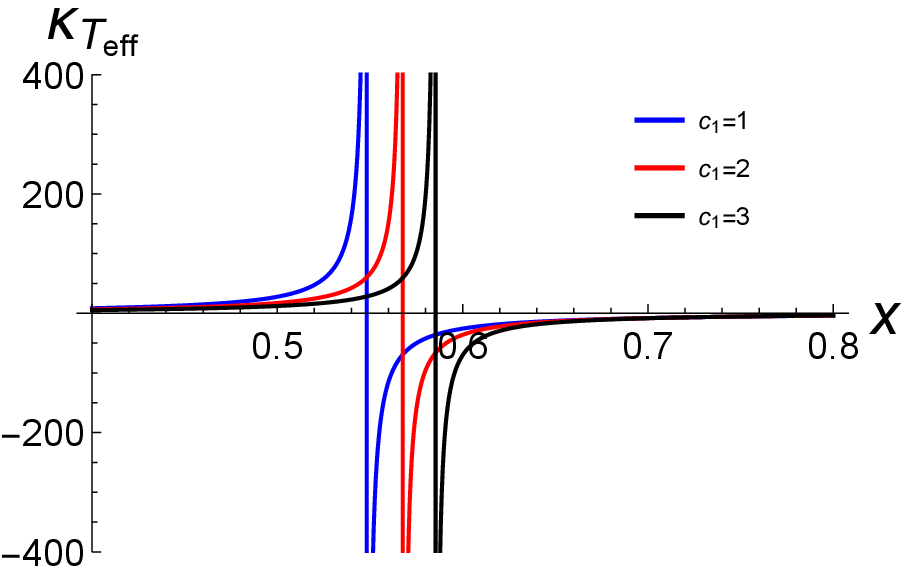}} %
\subfigure[]{\includegraphics[width=0.30\textwidth]{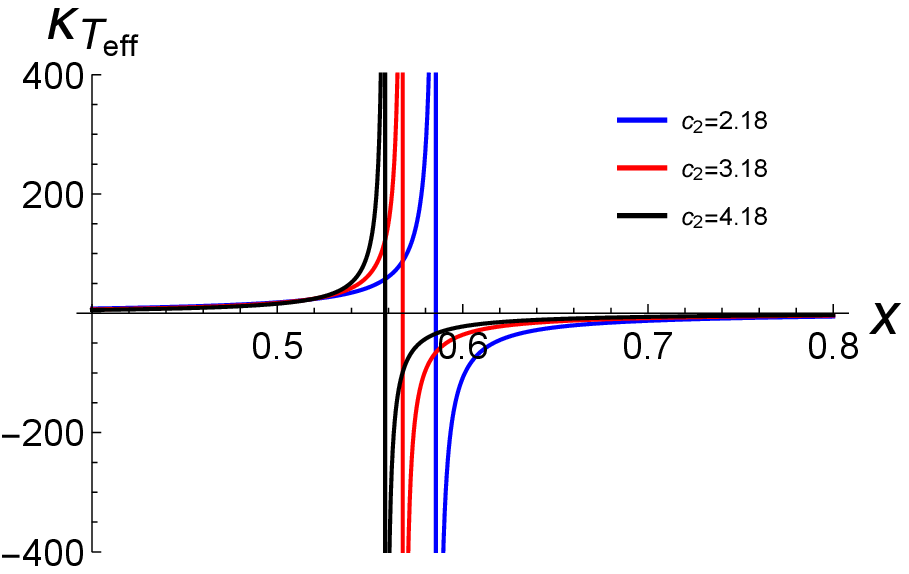}}\newline
\subfigure[]{\includegraphics[width=0.30\textwidth]{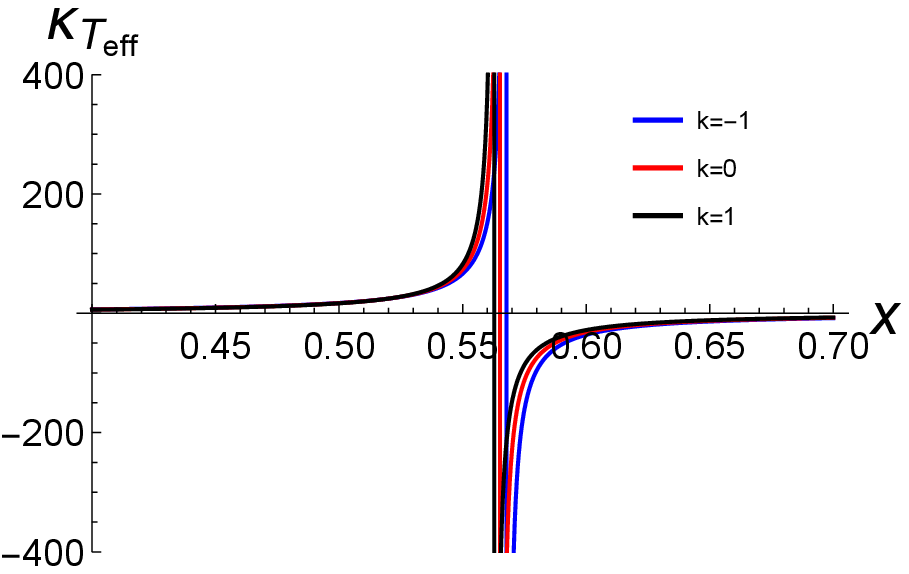}} %
\subfigure[]{\includegraphics[width=0.30\textwidth]{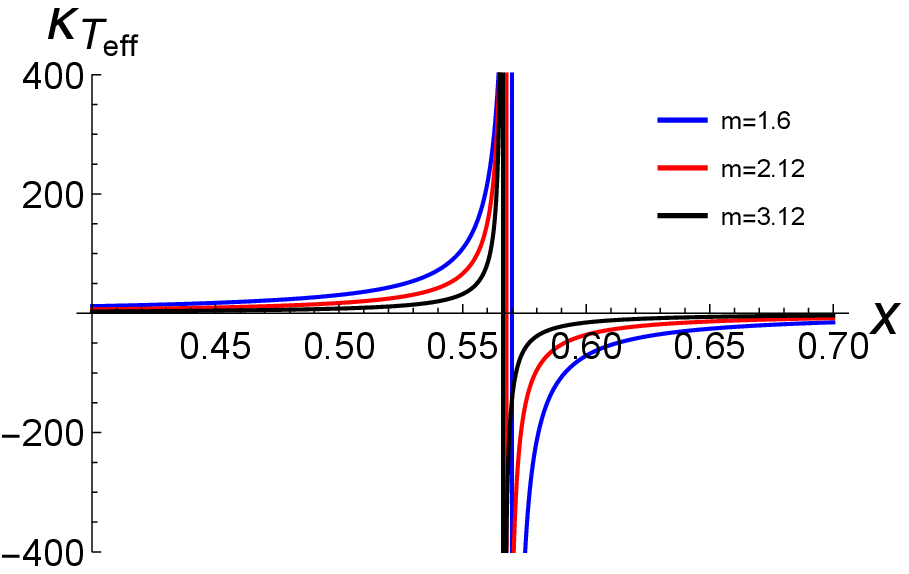}}
\subfigure[]{\includegraphics[width=0.30\textwidth]{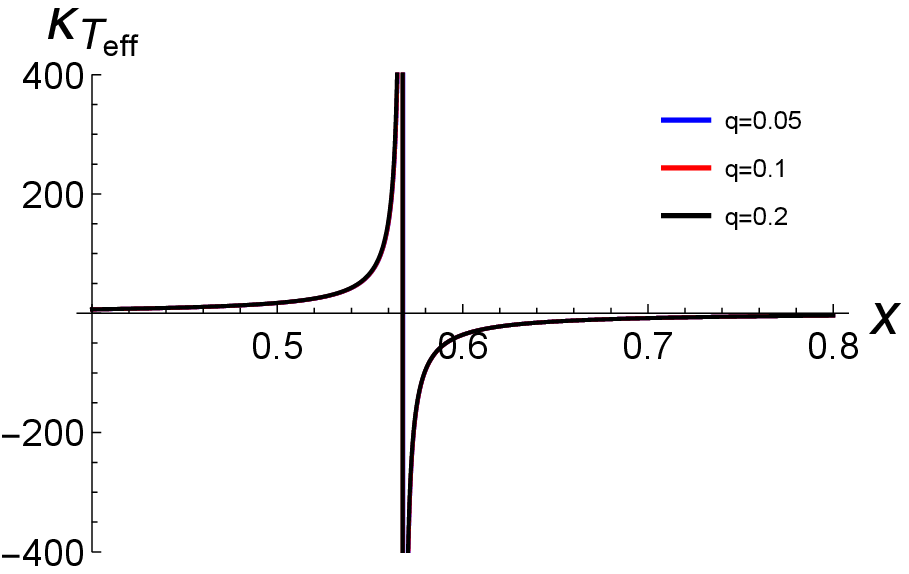}} %
\caption{$\kappa_{T_{eff}}-x$ diagrams when the parameters change respectively and the initial parameters  $m=2.12$, $c_1=2$,$c_2=3.18$,$q=1.7$, $r_c=1$,$k=1$.}
\label{fig6}
\end{figure}

According to  Eq.(\ref{3.6}), we plot $\kappa_{T_{eff}}-x$ curve in FIG.\ref{fig6}.

It can be seen from the curve $C_{P_{eff}}-x$, $\alpha-x$, $\kappa_{T_{eff}}-x$ , $C_{P_{eff}}$, $\alpha$ and $\kappa_{T_{eff}}$ are divergence at $x=x_c$. The entropy and volume of the system are continuous. According to Ehrenfest's classification of phase transition, $x_c$ is the second-order phase transition point of the system. 

 \begin{table}
 \end{table}

 Gibbs function is given by

\begin{equation}\label{3.7}
\begin{aligned}
G&=M-T_{eff} S\\
&=\frac{\left(k+m^{2} c_{2}\right) r_{c} x(1+x)}{2\left(1+x+x^{2}\right)}+\frac{q^{2}(1+x)\left(1+x^{2}\right)}{8 r_{c} x\left(1+x+x^{2}\right)}+\frac{r_{c}^{2} m^{2} c_{1} x^{2}}{4\left(1+x+x^{2}\right)}-\frac{r_{c} B(x, q)\left(1-x^{3}\right)}{4 x\left(1+x^{4}\right)}\left[1+x^{2}+f(x)\right].
\end{aligned}
\end{equation}

the $G-T_{eff}$ curve is following

\begin{figure}[htp]
\centering

 \subfigure[]{\includegraphics[width=0.30\textwidth]{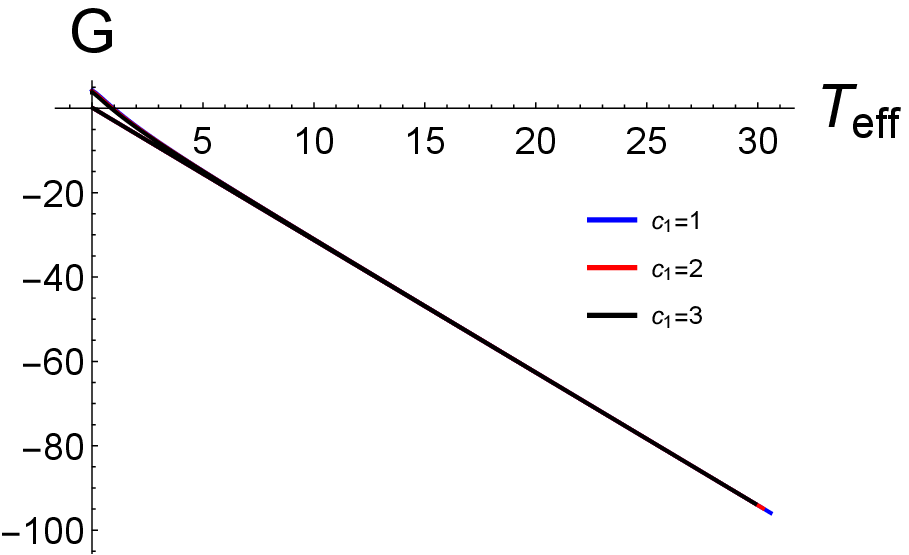}} %
\subfigure[]{\includegraphics[width=0.30\textwidth]{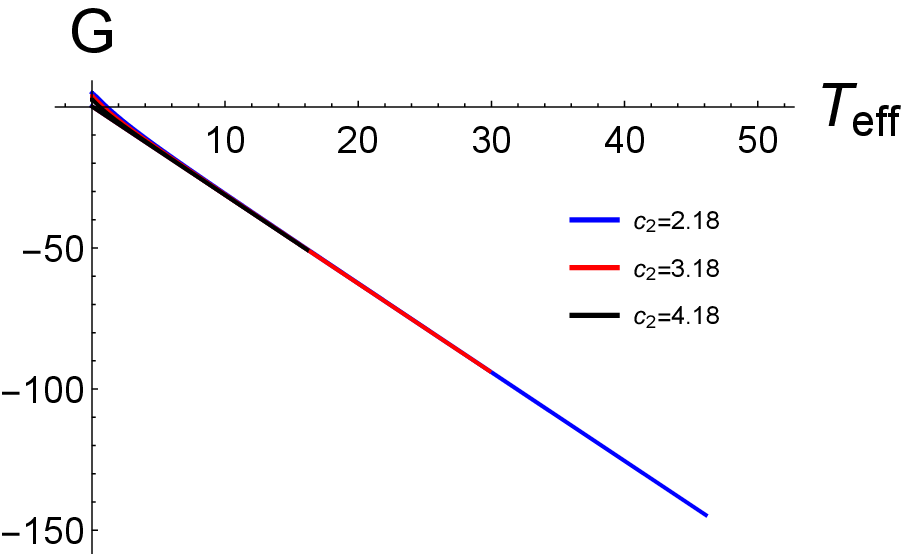}}\newline
\subfigure[]{\includegraphics[width=0.30\textwidth]{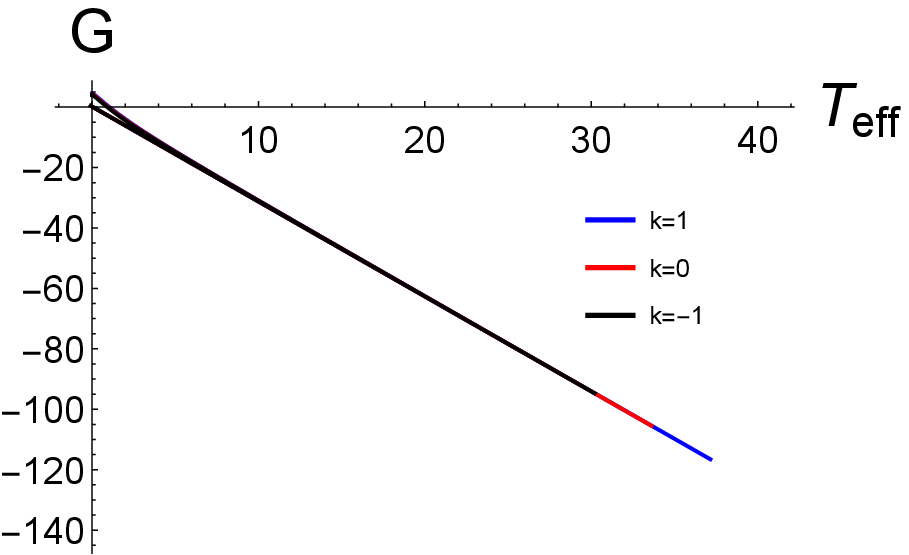}}
\subfigure[]{\includegraphics[width=0.30\textwidth]{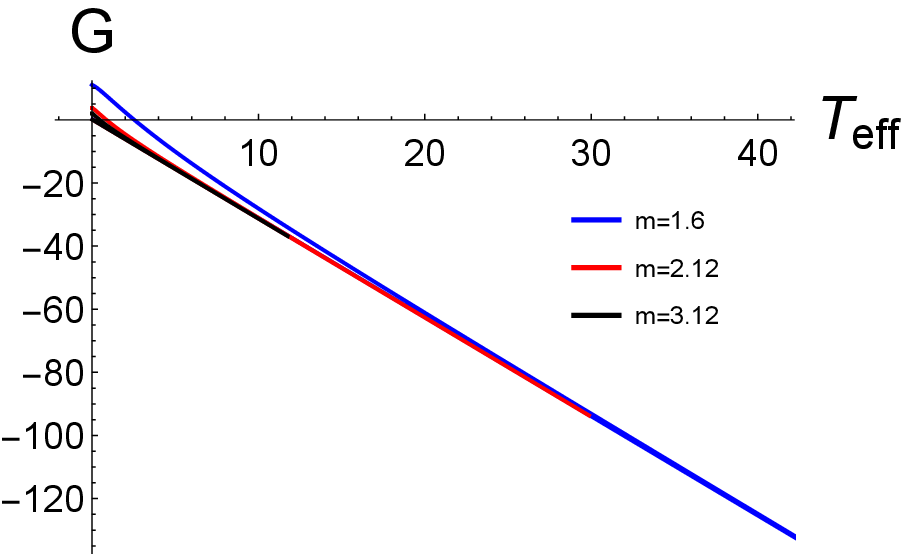}}
\subfigure[]{\includegraphics[width=0.30\textwidth]{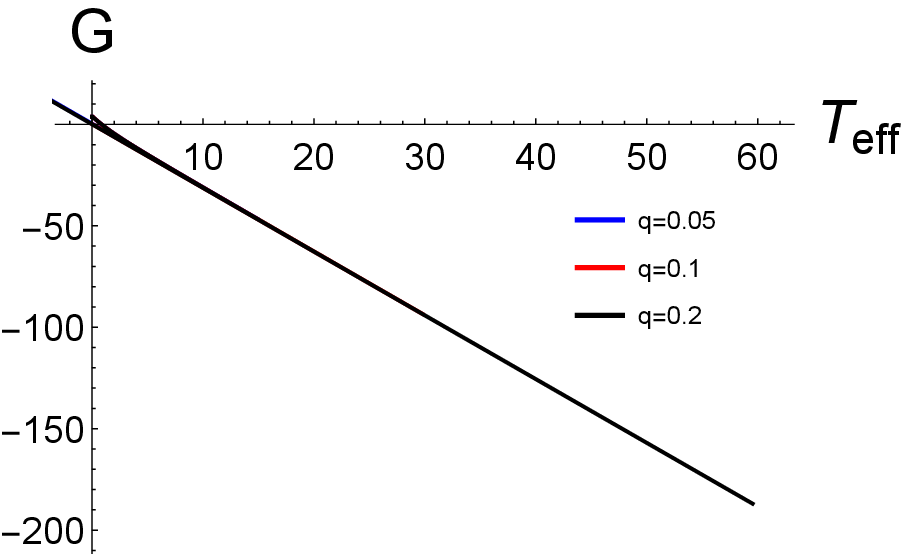}} %
\caption{$G-T_{eff}$ diagrams when the parameters change respectively and the initial parameters  $m=2.12$, $c_1=2$,$c_2=3.18$,$q=1.7$, $r_c=1$,$k=1$.}
\label{fig7}
\end{figure}
 From Figures $C_{P_{eff}}-x$, $\alpha-x$, $\kappa_{T_{eff}}$ and $G-T_{eff}$, DSBHMG satisfies Ehrenfest's second-order phase transition conditions for phase transition classification at $x=x_c$, so secondary phase transition occurs in DSBHMG space-time at $x=x_c$.

From the expressions (\ref{2.10}) and (\ref{2.12}) of the total entropy and volume of space-time, it is known that in the  range $0<x<1$ the entropy and volume of DSBHMG are continuous. According to the curve $G-T_{eff}$, DSBHMG  does not have a first-order phase transition, which is different from the first-order phase transition of AdS black holes.

\section{The entropic force of interaction between two horizons}

\label{sec:conclusion}

For a general thermodynamic system, when two systems are independent of each other, the corresponding entropy of system A and system B are $S_A$ and $S_B$. When two systems have interactions, the entropy of the total system is

\begin{equation}\label{4.1}
S_{total}=S_A+S_B+S_{AB}
\end{equation}

In the formula, the $S_{AB}$ is an extra entropy caused by the interaction of two systems.
From formula (\ref{2.12}), we know that the total entropy of the effective thermodynamic system in DSBHMG is divided into two parts, one is the entropy corresponding to the two horizons, and the other term is the increase of the system entropy after considering the interaction of the two horizons as a thermodynamic system, so the entropy is

\begin{equation}\label{4.2}
S_{f}=S_{A B}=\pi r_{c}^{2} f(x)=\pi r_{c}^{2}\left[\frac{8}{5}\left(1-x^{3}\right)^{2 / 3}-\frac{2\left(4-5 x^{3}-x^{5}\right)}{5\left(1-x^{3}\right)}\right]
\end{equation}

Entropic force in thermodynamic system is \citep{Erik P. Verlinde11,Plastino18a,Plastino18b,Li-Chun Zhang19,Panos19,Dmitri E. Kharzeev14}

\begin{equation}\label{4.3}
F=-T \frac{\partial S}{\partial r}
\end{equation}

which $T$ is the temperature of the system,  $r$ is the  location of the boundary surface.

The entropic force in DSBHMG is

\begin{equation}\label{4.4}
F=T_{eff}\left(\frac{\partial S_{f}}{\partial r}\right)_{T_{eff}}
\end{equation}

Where $T_{eff}$ is  the equivalent temperature of the system, $r=r_c-r_+=r_c(1-x)$, from Eq.(\ref{4.2}), we can obtain Eq.(\ref{4.5}),

\begin{equation}\label{4.5}
\begin{aligned}
 F(x)&=T_{eff} \frac{\left(\frac{\partial S_{f}}{\partial r_{c}}\right)_{x}\left(\frac{\partial T_{eff}}{\partial x}\right)_{r_{c}}-\left(\frac{\partial S_{f}}{\partial x}\right)_{r_{c}}\left(\frac{\partial T_{eff}}{\partial r_{c}}\right)}{(1-x)\left(\frac{\partial T_{eff}}{\partial x}\right)_{r_{c}}+r_{c}\left(\frac{\partial T_{eff}}{\partial r_{c}}\right)_{x}} \\
&=\frac{B(x, q)\left(1-x^{3}\right)}{4 x\left(1+x^{4}\right)} \frac{\left[2 B'(x, q)\left(1-x^{3}\right) f(x)-\frac{2 B(x, q)\left(1+2 x^{3}+5 x^{4}-2 x^{7}\right)}{x\left(1+x^{4}\right)} f(x)+\bar{B}(x, q)\left(1-x^{3}\right) f'(x)\right]}{\left[B'(x, q)\left(1-x^{3}\right)(1-x)-\frac{ B(x, q)\left(1+2 x^{3}+5 x^{4}-2 x^{7}\right)}{x\left(1+x^{4}\right)}(1-x)-\bar{B}(x, q)\left(1-x^{3}\right)\right]}.
\end{aligned}
\end{equation}

\begin{figure}[htp]
\subfigure[]{\includegraphics[width=0.30\textwidth]{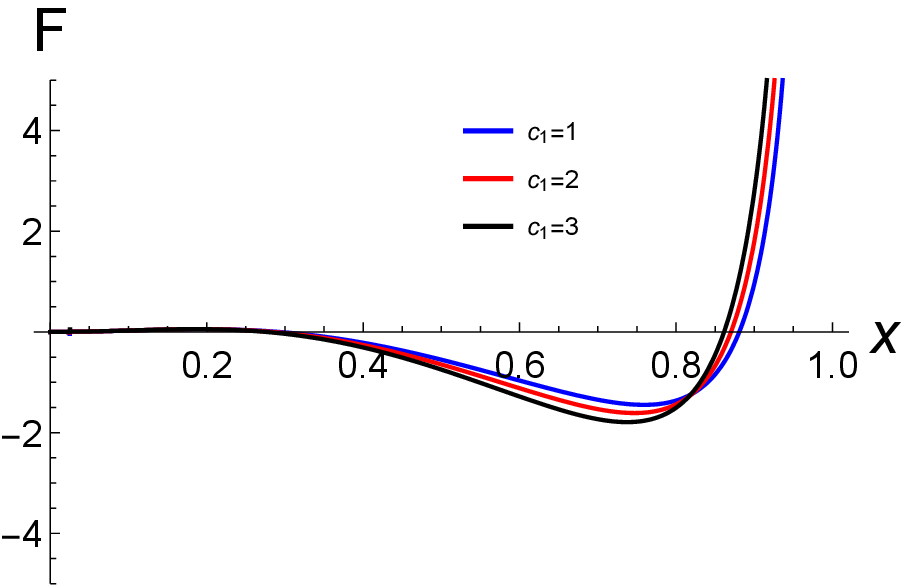}} %
\subfigure[]{\includegraphics[width=0.30\textwidth]{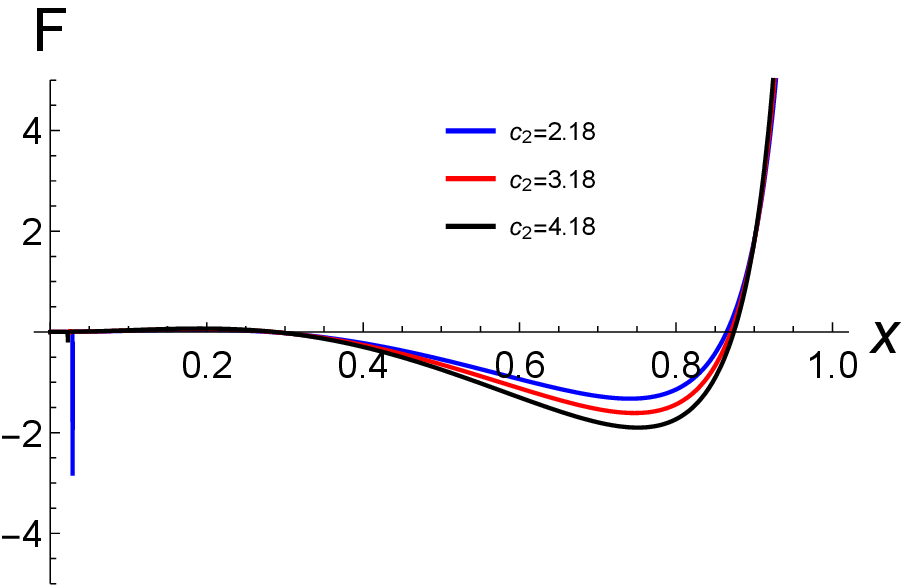}}\newline
\subfigure[]{\includegraphics[width=0.30\textwidth]{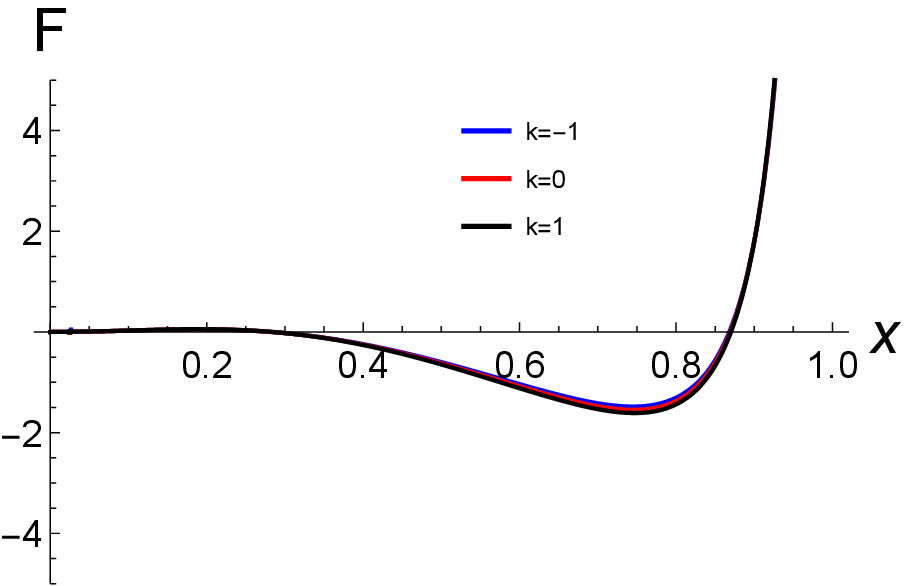}} %
\subfigure[]{\includegraphics[width=0.30\textwidth]{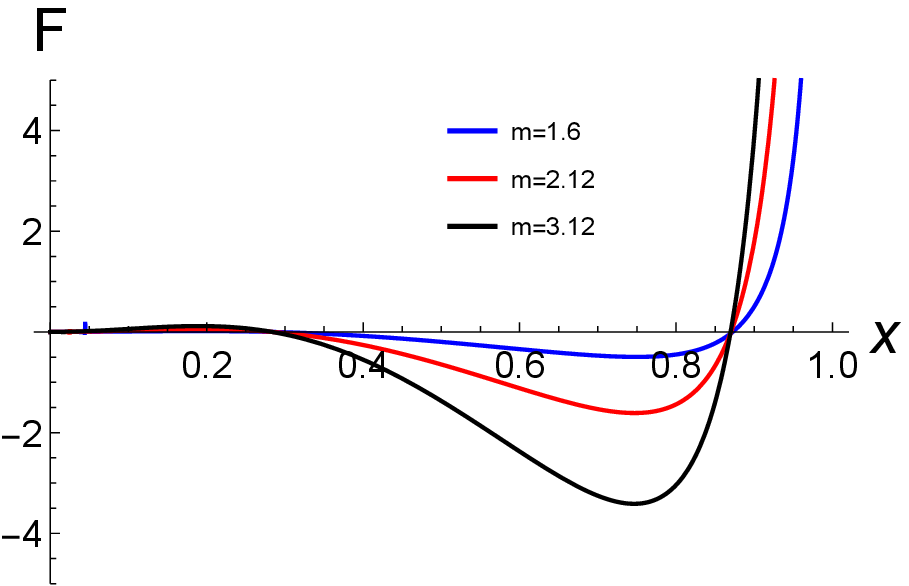}}
\subfigure[]{\includegraphics[width=0.30\textwidth]{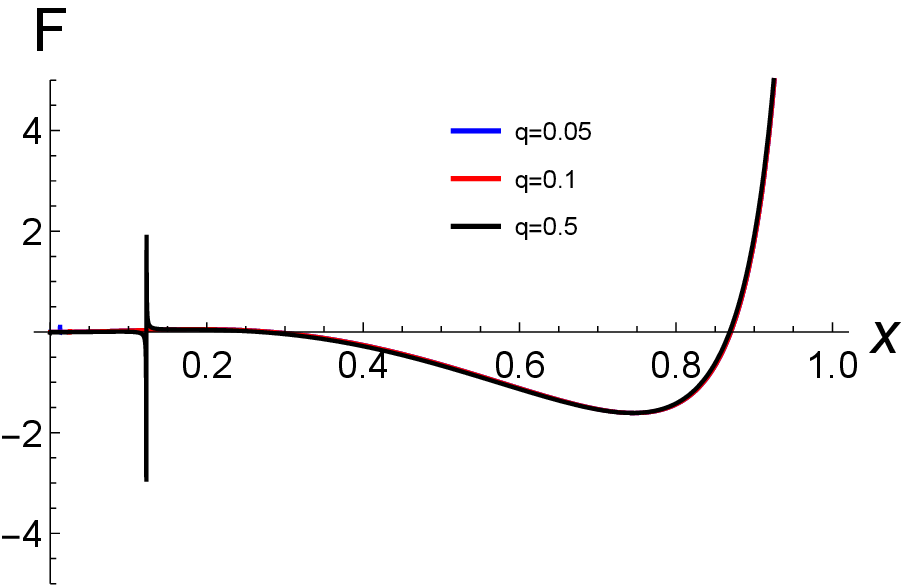}} %
\caption{$F-x$ diagrams when the parameters change respectively and the initial parameters  $m=2.12$, $c_1=2$,$c_2=3.18$,$q=1.7$, $r_c=1$,$k=1$.}
\label{fig8}
\end{figure}

The $F(x)-x$ curve is shown in FIG.\ref{fig8}.

For different parameters, the intersection position of the $F(x)-x$ curve and the axis is $x_0$. The entropic force between the two horizons is positive($F>0$) when $1>x>x_0$, which means that the two horizons are mutually exclusive; and the entropic force between the two horizons is negative ($F<0$) when $x_0>x>0$, indicating that the two horizons are mutually attractive. Therefore, the position ratio of the two horizons $x$ is different, and the entropic force between them is different. The two horizons are accelerated to separate under the action of entropic force, and the cosmological horizon expands faster than the black hole horizon when $1>x>x_0$, and they are decelerating to be separated by the entropic force, and the cosmological horizon undergoes a decelerating expansion relative to the black hole horizon when $x_0>x>0$. So we get that the different values $S_1-s_2$ between the area $S_1$ of the curve and the axis in interval $1>x>x_0$ and the area $S_2$ of the curve and the axis in interval $x_0>x>0$ determines whether the cosmological horizon accelerates or oscillates with respect to the black hole horizon. The cosmological horizon changes from accelerating expansion to decelerating expansion when $S_1-S_2>0$. When it is equal to or less than zero, the cosmological horizon is changing from accelerating expansion to decelerating expansion, and then changing from accelerating contraction to decelerating contraction. One cycle ends and the next cycle begins. The cosmological horizon is oscillating relative to the horizon of the black hole.

Interactions between neutral molecules or atoms with a center of mass separation $r$ are often approximated by the so-called Lennard-Jones potential energy $\phi_{L,J}$ which is given by \citep{David C. Johnston,Plastino18a,Plastino18b,Yan-Gang Miao18}
\begin{equation}\label{4.6}
\phi_{L, J}=4 \phi_{\min }\left[\left(\frac{r_{0}}{r}\right)^{12}-\left(\frac{r_{0}}{r}\right)^{6}\right],
\end{equation}
where the first term is a short-range repulsive interaction and the second term is a longer-range attractive interaction. A plot of $\frac{\phi_{L,J}}{\phi_{min}}$ versus $\frac{r}{r_0}$ is shown in FIG.\ref{fig3}. The value $r=r_0$ corresponds to $\phi_{L,J}=0$, and the minimum value of $\phi_{L,J}$ is $\frac{\phi_{L,J}}{\phi_{min}}=-1$ at

\begin{equation}\label{4.7}
\frac {r_{min}}{r_{0}}=2^\frac{1}{6}\approx 1.122
\end{equation}

By the definition of potential energy, the force between a molecule and a neighbor in the radial direction from the first molecule is $F_r=\frac{-d\phi_{L,J}}{dr}$ , which is positive (repulsive) for $r<r_{min}$ and negative (attractive) for $r>r_{min}$. When the center of the first particle coincides with the coordinate center, let the radius of the particle be $\frac{r_0}{2}$. When the center of the second particle is at $r$, and the boundary of the second particle is at $r_2$, then $r =r_2+ \frac{r_0}{2}$. Let's take $y=\frac{r_0}{2r_2}$, $0<y\leq1$. Eq.(\ref{4.6}) can be expressed as

\begin{equation}\label{4.8}
\phi_{L, J}(y)=4 \phi_{\min }\left[\left(\frac{r_{0}}{r}\right)^{12}-\left(\frac{r_{0}}{r}\right)^{6}\right]=4 \phi_{min} 2^{6}\left[2^{6}\left(\frac{y}{1+y}\right)^{12}-\left(\frac{y}{1+y}\right)^{6}\right].
\end{equation}

The interaction between the two particles is

\begin{equation}\label{4.9}
F(y)=-\frac{d \phi_{L, J}}{d r}=4 \phi_{\min } \frac{6}{r}\left[2\left(\frac{r_{0}}{r}\right)^{12}-\left(\frac{r_{0}}{r}\right)^{6}\right]=\frac{3 \phi_{min} 2^{10}}{r_{0}}\left[2^{7}\left(\frac{y}{1+y}\right)^{13}-\left(\frac{y}{1+y}\right)^{7}\right].
\end{equation}
\begin{figure}[htbp]
\centering
\begin{minipage}[t]{0.48\textwidth}
\centering
\includegraphics[width=3.3in]{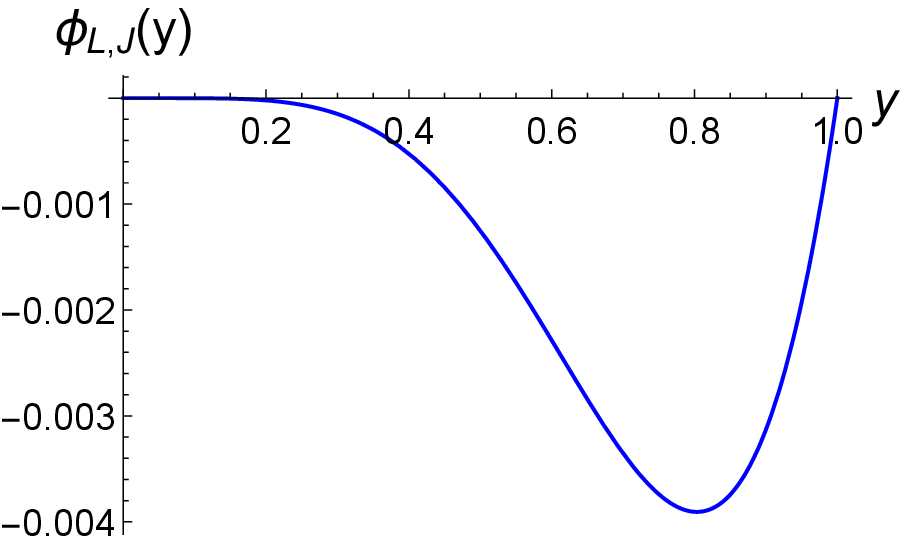}
\caption{$\phi_{L, J}(y)-y$ diagram when $y$ from $0$ to $1$.}
\label{fig9}
\end{minipage}
\begin{minipage}[t]{0.48\textwidth}
\centering
\includegraphics[width=3in]{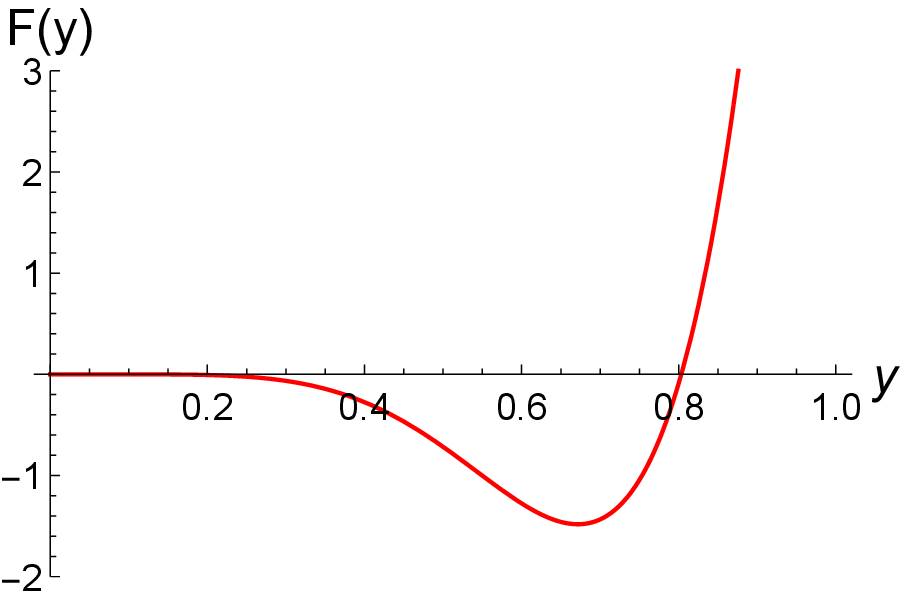}
\caption{$F(y)-y$ diagram when $y$ from $0$ to $1$.}
\label{fig10}
\end{minipage}
\end{figure}

From FIG.\ref{fig8} and FIG.\ref{fig10}, we can see the relation curve of entropic force between two horizons in DSBHMG with the ratio of the position of the two horizons is very similar to that of the Lenard-Jones interaction force with the position ratio of the two particle boundaries, which has the same change rule.

\begin{figure}[htp]
\centering
\subfigure[]{\includegraphics[width=0.45\textwidth]{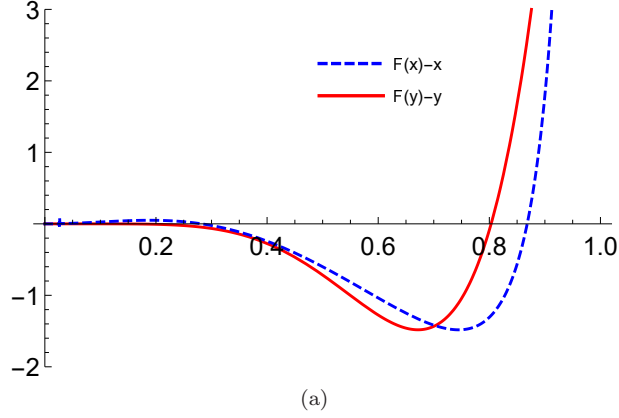}}
\caption{$F(x)-x$ and $F(y)-y$ diagrams when abscissa $x$ and $y$ from $0$ to $1$.}
\label{fig11}
\end{figure}

In order to show the similarity of the two curves more clearly, we show the two curves in the same coordinate as is shown in FIG.\ref{fig11}.

From FIG.\ref{fig11}, we know that the trend of change in curve $F(x)-x$ and $F(y)-y$ is exactly the same, only the intersection position of the two curves and the abscissa is slightly different. It shows that the entropic force between the horizon of black hole and the cosmological horizon is very similar to the Lenard Jones interaction force between two particles in RN dssq space-time, which provides a new way to explore the interaction force between particles in black hole.

\section{Conclusion and discussion}

When the black hole horizon and the cosmological horizon are viewed separately as independent thermodynamic systems without considering the correlation between them, the space-time does not meet the requirements of thermodynamic equilibrium stability because the radiation temperature of the two horizons are different. Thus the space-time is unstable. When the correlation of two horizons is considered, the effective temperature $T_{eff}$, effective pressure $P_{eff}$ and effective potential $\phi_{eff}$ reflecting the thermal properties of DSBHMG are given by Eq.(\ref{2.14}). From curve $C_{p_{eff}}-x$, $\alpha-x$ and $\kappa_{T_{eff}}-x$, when the ratio of the positions of the two horizons in DSBHMG is $x=x_c$, phase transition occurs in the system. According to Ehrenfest's classification of phase transition, the phase transition of the system at this point is of second-order, which is similar to that of AdS black hole. In the study of the thermodynamic properties of spherically symmetric AdS black holes, people study the critical phenomena of AdS black holes by comparing the thermodynamic quantities of AdS black holes with the thermodynamic quantities of Van der Waals equation and obtain the critical exponents and the heat capacity of AdS black holes, which provides a basis for further study of the thermal effect of black holes and experimental observation. However, it is difficult to accept the fact that the heat capacity at constant volume of AdS black holes is zero. It is shown that the effective thermodynamic quantity of DSBHMG also has the phase transition characteristics similar to that of van der Waals system. From formula (\ref{3.1}), the heat capacity of DSBHMG at constant volume is not zero, which is consistent with the result that this heat capacity of van der Waals system is not zero. From the curve of $C_{P_{eff}}-x$, we know that the stability of DSBHMG depends on the value of  $x$ . When $x<x_{c}$, space-time meets the requirements of thermodynamic stability. When $1>x>x_{c}$, space-time does not meet the requirements of thermodynamic equilibrium stability, and space-time is unstable. Therefore, there is no DSBHMG steady-state space-time satisfying the position ratio $1>x>x_{c}$ of two horizons in the universe, which provides a theoretical basis  to find black holes. The influence of parameters in DSBHMG on space-time stability is shown in $C_{P_{eff}}-x$ curve seeing FIG.\ref{fig4}.

In the framework of general relativity, the entropic force of interaction between the horizon of black hole and the cosmological horizon given by theory is very similar to the Lennard-Jones force between two particles confirmed by experiments. Since the space-time metric is derived from relativity, the thermodynamics of space-time is based on general relativity, and the thermal effect of black hole is obtained from quantum mechanics. The physical quantities obtained meet the first law of thermodynamics. Therefore, the conclusion we have given reveals the internal relationship among general relativity, quantum mechanics and thermodynamics, and provide a new way for us to study the interaction between particles in black holes and the micro state of particles in black holes, and the relationship between Lennard-Jones potential and micro state of particles in ordinary thermodynamic systems.

Through the analysis of the fourth part, it is known that under the action of entropic force, when the ratio of the positions of the two horizons satisfies $x_0<x<1$, the entropic force between the two horizons is mutually exclusive, and the two horizons in this region are accelerating expansion under the action of entropic force; when $0<x<x_0$, the entropic force between the two horizons is mutually attractive, and the two horizons are decelerating expansion; until $x\rightarrow{0}$, the entropic force between the two horizons is zero, and the two horizons tend to be relatively static. Therefore, in the $0<x<1$ range, the cosmological horizon accelerates to expand relative to the horizon of the black hole under the action of the entropic force. Our universe is a quasi de Sitter space-time, so our discussion provides a new way to explore the internal cause of the expansion of the universe.

\section*{Acknowledgments}

We thank Prof. Z. H. Zhu for useful discussions.

This work was supported by the Scientific and Technological Innovation Programs of Higher Education Institutions of Shanxi Province, China (Grant No. 2020L0471, No. 2020L0472, No. 2019L0743) and the National Natural Science Foundation of China (Grant Nos. 12075143, 11847123, 11475108, 11705106, 11705107, 11605107).

\end{document}